\documentclass[preprint,nofootinbib,tightenlines,eqsecnum,floats,aps,amsfonts]{revtex4}

\usepackage{amsmath, graphicx, graphics, color}
\newcommand{\Lie}[0]{{\cal L}\, }

\newcommand{\tl}{\theta_{(\ell)}}
\newcommand{\tn}{\theta_{(n)}}

\newcommand{\nn}{\nonumber}
\newcommand{\be}{\begin{equation}}
\newcommand{\ee}{\end{equation}}
\newcommand{\bea}{\begin{eqnarray}}
\newcommand{\eea}{\end{eqnarray}}

\newcommand{\tq}{\tilde{q}}
\newcommand{\hu}{\hat{u}}
\newcommand{\hn}{\hat{n}}

\newcommand{\bfl}{\mbox{\boldmath{$\ell$}}}
\newcommand{\bfn}{\mbox{\boldmath{$n$}}}
\newcommand{\bta}{\mbox{\boldmath{$\tau$}}}
\newcommand{\bxi}{\mbox{\boldmath{$\xi$}}}

\newcommand{\bu}{\bar{u}}
\newcommand{\bn}{\bar{n}}
\newcommand{\htau}{\hat{\tau}}
\newcommand{\hr}{\hat{r}}

\newcommand{\bV}{\bar{V}}
\newcommand{\bN}{\bar{N}}

\newcommand{\tom}{\tilde{\omega}}
\newcommand{\tx}{\tilde{x}}

\newcommand{\vM}{\mbox{\boldmath$\epsilon$}}
\newcommand{\vB}{{}^{{}^{(B)}} \! \! \mbox{\boldmath$\epsilon$}}
\newcommand{\vSigma}{{}^{{}^{(\Sigma)}} \! \! \mbox{\boldmath$\epsilon$}}
\newcommand{\vBt}{\bar{\mbox{\boldmath$\epsilon$}}}
\newcommand{\vH}{{}^{{}^{(H)}} \! \! \mbox{\boldmath$\epsilon$}}

\newcommand{\vS}{\widetilde{\mbox{\boldmath$\epsilon$}}}

\newcommand{\epSigma}{{}^{{}^{(\Sigma)}} \! \! \epsilon}

\newcommand{\bdv}{\mathbf{d}\mbox{\boldmath{$v$}}}

\newcommand{\lpb}{\underleftarrow{\mbox{\boldmath{$\ell$}}}}
\newcommand{\npb}{\underleftarrow{\mbox{\boldmath{$n$}}}}
\newcommand{\bP}{\mathbf{P}}
\newcommand{\bcH}{\mbox{\boldmath$\mathcal{H}$}}

\newcommand{\bL}{\mbox{\boldmath$L$}}

\newcommand{\cV}{\mathcal{V}}

\begin{document}

\title{Horizon energy and angular momentum from a Hamiltonian perspective}

\author {Ivan Booth${}^{1,2}$}
\email{ibooth@math.mun.ca}
\author{Stephen Fairhurst$^{1,3}$}
\email{sfairhur@gravity.phys.uwm.edu}
\address{1. Theoretical Physics Institute, Department of Physics, 
            University of Alberta, Edmonton, Alberta, T6G 2J1, Canada \\
         2. Department of Mathematics and Statistics,
            Memorial University of Newfoundland,
            St. John's, Newfoundland and Labrador, A1C 5S7, Canada \\
         3. Department of Physics, 
            University of Wisconsin -- Milwaukee,
            P.O. Box 413, Wisconsin, 53201, U.S.A.}

\begin{abstract} 
Classical black holes and event horizons are highly non-local objects, defined
in terms of the causal past of future null infinity.  Alternative,
(quasi)local definitions are often used in mathematical, quantum, and
numerical relativity.  These include apparent, trapping, isolated, and
dynamical horizons, all of which are closely associated to two-surfaces of
zero outward null expansion. In this paper we show that three-surfaces which
can be foliated with such two-surfaces are suitable boundaries in both a
quasilocal action and a phase space formulation of general relativity. The
resulting formalism provides expressions for the quasilocal energy and angular
momentum associated with the horizon.  The values of the energy and angular
momentum are in agreement with those derived from the isolated and dynamical
horizon frameworks.  
\end{abstract}

\maketitle

%%%%%%%%%%%%%%%%%%%%%%%%%%%%%
% INTRODUCTION
%%%%%%%%%%%%%%%%%%%%%%%%%%%%%

\section{Introduction}

The basic ideas of black hole physics are well-known. These foundations,
including the singularity, uniqueness, and area increase theorems
\cite{hawkellis}, the laws of black hole mechanics \cite{bch}, and the identity
between those laws and the laws of thermodynamics \cite{bek, hawk}, have been
understood for several decades. Perhaps the most remarkable conclusion to be
garnered from these results is the realization that black holes, in spite of
their apparently exotic origin, are not really so different from other objects
studied by physics.  Though classically they harbour singularities in spacetime
and have interiors that cannot causally effect the rest of the universe, they
still have physical properties such as mass, charge, angular momentum,
temperature, and entropy. Nevertheless there is at least one fundamental way in
which they are very different.  With the exception of spacetimes that are known
to be globally stationary, the exact location and size of a black hole cannot
be determined by local measurements; instead the existence of a black hole
event horizon is a property of the causal structure of the spacetime. Further,
in traditional black hole physics, physical properties such as mass cannot be
determined by local measurements, but instead are only defined with reference
to asymptotic structures. This, of course, contrasts quite strongly with most
physical objects which can be identified, measured, and understood without
reference to the far future and spatial infinity.

Given the non-local nature of event horizons, there has always been an
interest in alternative, local, characterizations of black holes. 
The original alternative black hole signature was the trapped 
surface \cite{penrose} and these figure prominently in the definition of
apparent horizons \cite{hawkellis}. More recently there have been several 
other (closely related) approaches. These
include trapping \cite{hayward, hay2}, isolated \cite{ih}, dynamical
\cite{ak}, and slowly evolving horizons \cite{prl}.  All of these will be
discussed in more detail in Section \ref{horizons} but for now the important
point to note is that they are all closely related to each other and defined
by the convergence/divergence of particular sets of null geodesics.
Specifically they all share one common property : the three-surface which
they identify as a ``horizon" can be foliated with spacelike two-surfaces of
vanishing outward null expansion.  These alternative notions of horizons are
especially suited to studies of dynamic black holes.  For an evolving black
hole, the location of an event horizon is impossible to determine unless one
first understands the causal structure of the entire spacetime. The
alternative horizons do not face this problem as their location depends only
on local geometry --- the calculations involved may be difficult, but at least
they are possible. 

Thus, alternative horizons are closely associated with the study of black hole
evolutions and their interactions with their surroundings. The exchange of
energy and angular momentum between a black hole and its environment is one of
the most basic physical problems that might be investigated under these
circumstances and this question has been addressed under each of the
approaches. Various formulae have been proposed for energy and
angular momentum associated with a horizon, as well as the flux of these
quantities into and out of the hole. 

In this paper, we apply Hamiltonian methods to gain further insight into the 
energy and angular momentum of horizons. Specifically, we examine quasilocal
action and Hamiltonian formulations of general relativity on manifolds which
have horizons as boundaries.  We begin in Section \ref{actSect} with a
discussion of the action in the spirit of the well-known Brown--York
quasilocal action formalism \cite{by1,by2} and its non-orthogonal
generalizations \cite{nopaper, BLY}.  We show that any three-surface
(regardless of signature) which can be foliated with spacelike two-surfaces of
vanishing outward null expansion is a suitable boundary in a quasilocal action
formulation for general relativity, provided that we fix the intrinsic geometry
of those two-surfaces along with the connection on their normal bundles (a
quantity closely associated with angular momentum).  This boundary information
only needs to be fixed in a weak form --- up to diffeomorphisms which may
``move it around''.  

If we impose a notion of time on the quasilocal region, then we can perform a
Legendre transform of the action and so obtain a Hamiltonian for the
spacetime.  As in other derivations of a Hamiltonian, the bulk part of this
functional is made up of constraints and so vanishes on solutions of the
Einstein equations.  Thus one can evaluate it on a slice of a solution by
examining only the geometry of the boundary.  Consequently, we can derive a
quasilocal energy expression for these alternative horizons.  As usual, the
value of the Hamiltonian is sensitive to the precise form of the action.  With
our boundary conditions, there is a class of acceptable action functionals;
the freedom in the action amounts to being allowed to add free functionals of
the fixed data to the original form.  As a consequence, the formalism fixes
the energy only up to free functionals of the fixed data. 

In Section \ref{cpsf}, we turn from Lagrangian to Hamiltonian arguments and
consider the horizons as boundaries from the point of view of phase space, a
symplectic structure, and Hamiltonian evolutions of a quasilocal region of
spacetime. This investigation helps to further elucidate the allowed freedom in
the definition of an energy and also gives us a concrete expression for the
angular momentum associated with the horizon. The calculations follow the
extended phase space formalism of \cite{bf1} which is designed for situations
where there are fluxes of quantities such as energy and angular momentum
through the boundaries. This is in contrast to the standard Hamiltonian
formalism which is designed for isolated systems where there can be no flux of
these quantities (and so they are conserved).  

Applying this analysis, we derive expressions for the energy and the angular
momentum associated with the horizon.  The energy expression so derived
matches the one found using the action/Legendre transform methods --
essentially any diffeomorphism invariant functional of the intrinsic boundary
two-geometry and the connection on its normal bundle. The new piece of
information is the expression for the angular momentum.  

Finally, in Section \ref{energy} we compare the Hamiltonian expressions for
energy and angular momentum with those arising from the isolated and dynamical
horizon calculations.  On sections of the horizon which are null, we find that
the angular momentum is identical to that obtained in the isolated horizon
framework.  Additionally, if we restrict our generators of rotation to divergence 
free vector fields, the associated angular momenta
on spacelike sections of the horizon agree with
the dynamical horizon values.  Next, we turn to the energy.  We can greatly
restrict the freedom in the energy by requiring that it satisfy a basic
property: the value of the energy must transform appropriately with rescalings
of the evolution vector field.  We find that in this case the natural evolution
vector field is the null vector, $\ell$, for which the outward expansion
vanishes. This is true even though it is not, in general, tangent to the
horizon.  Furthermore, the energy associated to a given evolution is then
determined up to an overall constant rescaling.  This constant can be fixed by
appealing to the dynamical horizon framework.  With this choice, the energy
derived using Hamiltonian methods will satisfy the dynamical horizon flux law.

%%%%%%%%%%%%%%%%%%%%%%%%%%%%%%%%
% HORIZONS
%%%%%%%%%%%%%%%%%%%%%%%%%%%%%%%%

\section{Local Horizons - a review}

\subsection{Definitions}
\label{horizons}

There are several (quasi-)local definitions of a horizon.  In this section, we
summarize the various definitions, highlighting both their similarities and
differences, and emphasizing those features relevant for the action and
Hamiltonian calculations that will follow.  

Spacelike two-surfaces and their null normal vectors feature prominently in
each of the alternative horizon definitions, so we begin by discussing them.
For any spacelike two-surface $S$, we can find families of null vectors that
are normal to the surfaces. Up to normalization and orientation, there are just
two such null directions. If the spacetime is time-oriented we can single out
the forward-in-time pointing directions and so, up to rescaling, define two
future-pointing null normals $\ell^a$ and $n^a$. Then, if $\tq_{ab}$ is the
two-metric on $S$, the expansion associated with these vector fields is given
by:
\begin{eqnarray} \tl = \tq^{ab} \nabla_{a} \ell_{b} \mspace{4mu} \mbox{ and }
\mspace{4mu} \tn = \tq^{ab} \nabla_{a} n_{b} \end{eqnarray}
These are expansions in the sense that if one Lie-drags $S$ in the direction
$\ell^a$, and $\vS$ is the area two-form on $S$, then 
\begin{eqnarray} \Lie_\ell \vS = \vS \tl \, .  \end{eqnarray}
Thus, $\tl$ describes how the area-form on $S$ expands/contracts along $\ell$.
Similar arguments apply to $\tn$. 

The two-surface $S$ is said to be trapped if it is closed and the expansions of both
$\ell$ and $n$ are negative at every point on $S$. A marginally trapped
surface is a slight generalization of a trapped surface; in this case, the
expansions of the null vectors are required to be less than \textit{or equal}
to zero.  Trapped and marginally trapped surfaces can be contrasted to
surfaces in flat spacetime where one would expect the outward expansion of
convex surfaces to be positive everywhere, and the inward expansion negative.
Intuitively, the negative expansion of both null normals can be attributed to
strong gravitational fields in the interior of $S$.  This picture is
strengthened by the singularity theorems which state that (provided energy
conditions are satisfied), any spacetime containing a trapped surface must
also contain a singularity \cite{hawkellis, wald}.  Furthermore, it can be
shown that any trapped surface in an asymptotically flat spacetime must be
contained within an event horizon. Even in the absence of an appropriate asymptotic
structure, a trapped surface may always be extended into a spacelike three-surface that 
hides its inside from its outside \cite{israel}.  Thus, it seems natural that the various
notions of a horizon discussed below all make use of trapped or marginally
trapped surfaces.

We begin with a brief discussion of apparent horizons \cite{hawkellis}.
Let $\Sigma$ be an asymptotically flat Cauchy surface. 
From an intuitive point of view, such a
$\Sigma$ is an ``instant'' in an asymptotically flat spacetime.  Now, a
point $p \in \Sigma$ is said to be trapped if it lies on a trapped surface that in 
turn is contained entirely in $\Sigma$;
%three-dimensional region $C$ of $\Sigma$ is said to be a trapped region if it
%has a smooth two-boundary $\partial C$ over which $\tl \leq 0$ everywhere,
%where $\ell^a$ is chosen as the out-of-$C$, towards ``infinity", pointing null
%normal.  
the total trapped region $\mathcal{T}$ is the union of all the
trapped points in $\Sigma$.  An apparent horizon is then
defined as the boundary $\mathcal{A}$ of the total trapped region
$\mathcal{T}$.  With certain assumptions about smoothness \cite{kriele}, it can be shown, as
one would expect, that $\tl = 0$ on $\mathcal{A}$ .  Further, it is true that
$\mathcal{T}$ is itself contained within an event horizon. In the case of
stationary black hole solutions, such as Schwarzschild, $\mathcal{T}$
completely fills the region of $\Sigma$ inside the black hole and the apparent
horizon coincides with the event horizon. 

While it is certainly possible to identify an apparent horizon without
reference to the far future, it is not a trivial task to first locate all
the trapped surfaces, find their union and then obtain the boundary of that region.
Further, one requires knowledge of the entire ``instant'' before one can
locate the apparent horizon.  Finally, $\mathcal{A}$ won't always satisfy
the smoothness assumptions mentioned in the previous
paragraph (see \cite{chrusciel} for a good discussion). All of these factors
can make working with $\mathcal{A}$ inconvenient (at the very least).

More recent quasilocal horizon definitions attempt to extract the idea of
apparent horizons as the boundaries of regions of especially strong
gravitational fields, while dropping some of their more unwieldy aspects.
Specifically they circumvent a consideration of entire trapped regions and
focus directly on two-surfaces $S$ of vanishing outward null expansion $\tl =
0$.  This has the advantage of making the definitions quasilocal.
Furthermore, as is the case for apparent horizons, in asymptotically flat
spacetimes such surfaces $S$ are always contained within event horizons.  In
common usage, for example in numerical relativity \cite{thomas}, the term
``apparent horizon'' is frequently used to refer to the outermost two-surface
for which $\tl=0$ as it is simply not practical to find the boundary of the
total trapped region.  However, one should keep in mind that while in some
circumstances the definitions are equivalent, these surfaces do not always
coincide with $\mathcal{A}$. 

Notions such as trapping, isolated, and dynamical horizons seek to formalize
these vernacular ideas of an apparent horizon and at the same time remove their
somewhat problematic dependence on $\Sigma$ slicing.
\footnote{
For example, it is well-known that even Schwarzschild spacetime may
be foliated so that there are no apparent horizons in any slice \cite{wald1} .}  
Rather than considering two-dimensional surfaces with vanishing outward
expansion in some three-surface $\Sigma$, they instead define three-dimensional
horizons within the full four-dimensional spacetime.  These horizons are
required to be foliated by closed spacelike two-surfaces $S_v$ on which $\tl = 0$.  Thus, while
slicings of the spacetime may be found that are compatible with these $\tl = 0$
surfaces (in which case they will often be apparent horizons), the horizon
definitions are independent of any such foliation of the spacetime.\footnote{\label{two} 
Note however that
the slicing dependence discussed in the last footnote is then replaced by a non-rigidity of the horizons.  
In general the two-surfaces making up these horizons may be smoothly deformed while
keeping their $\tl = 0$ property. See \cite{gregabhay} for a recent discussion of this non-uniqueness.}

We will begin by considering isolated horizons and then move on to the very closely
related trapping and dynamical horizons. For detailed discussions of the merits
of each approach the reader is referred to the literature. Here we will briefly
review their relevant properties and demonstrate that each falls within the
scope of the upcoming calculations. 

Isolated horizons \cite{ih} are intended to characterize black holes that are
not interacting with their surroundings, even though those surroundings may
themselves be dynamic. An isolated horizon is a null three-surface whose null
normal has zero expansion.  Every foliation of an isolated horizon will have
the same outward null normal, whence the foliation into two-surfaces is not
important. Making use of a mild energy condition, it is straightforward to
demonstrate that isolated horizons don't change in either shape or size and no
energy flows across them.  With an additional restriction on the permitted
class of null normals $\ell^{a}$, it can be shown that isolated horizons obey
the zeroth law of black hole mechanics, namely that the surface gravity is
constant over the horizon.  Furthermore, a phase space version of the first law
has been shown to hold.  Finally, loop quantum gravity calculations have
demonstrated that they have an entropy proportional to their surface area
\cite{abk}.

Trapping horizons \cite{hayward} are defined as three-surfaces (of
indeterminate signature) that can be foliated with $\tl = 0$ two-surfaces $S$.
By themselves these conditions are not enough to distinguish between outer and
inner black hole horizons or for that matter between black holes and white
holes. Additional conditions are added to define future outer trapping horizons
(FOTHs).  A FOTH is a trapping horizon that also satisfies the property that
$\tn < 0$ and $\Lie_n \tl < 0$. Thus, ingoing null geodesics are converging,
while outgoing null geodesics converge inside the horizon, diverge outside and
have zero expansion on the horizon.  Given a three-surface $\Sigma$ that
intersects a FOTH along a foliating two-surface $S$, these conditions are
intended to ensure that the region inside of $S$ is trapped, while the region
outside will not be (thus $S$ should be apparent horizon).
In a spacetime respecting the null energy condition, the above
conditions are sufficient to guarantee that dynamic trapping horizons are
spacelike and expanding, and so individually obey a form of the second-law of
black hole mechanics.  In the absence of (gravitational wave) shearing or
matter falling across the horizon, they are null and non-expanding.  Finally
for energy condition violating matter, horizons can shrink in size, in accord
with Hawking radiation.

A slowly evolving horizon \cite{prl} is an ``almost" isolated FOTH and so
defined in the same way but with an extra condition that characterizes
``almost". Intuitively, these horizons can be thought of as FOTHs which are
only weakly interacting with their environment.  In this regime, a truly
dynamical version of the first law for black holes has been formulated.  This
is also the regime of interest for studying the approach to equilibrium after a
gravitational collapse or black hole merger.

A dynamical horizon (DH) \cite{ak} is also closely related to a FOTH. It is a
spacelike three-surface that can be foliated with spacelike two-surfaces for
which $\tn < 0$ and $\tl = 0$.  Here, the requirement that the surface be
spacelike has replaced the $\Lie_n \tl < 0$ condition of the FOTH. This
ensures that the horizon can be identified without the need to take
derivatives out of the three-surface.  In cases where the horizon is expanding
(which are largely the cases of interest) the Einstein equations along with
energy conditions ensure that the definitions of dynamical horizons and FOTHs
are equivalent. It has been shown that dynamical horizons satisfy a variety of
conservation laws that relate the change in horizon properties to fluxes of
stress-energy and matter across the horizon.  The connection between those
laws and the calculations of this paper will be discussed further in Section
\ref{energy}.

Finally, there has recently been some discussion of non-FOTH/DH extensions of
FOTHs/DHs \cite{bendov, gregabhay,bbgv}. Referred to as marginally trapped
tubes (MTTs), these are three-surfaces that satisfy $\tl = 0$ and $\tn < 0$ but
are not not necessarily everywhere null or spacelike. In the terminology of
trapping horizons, they are future trapping horizons with the slight
generalization that $\Lie_n \tl$ may take any value. It has been shown that a
single MTT can include not only regions that are isolated and dynamical but
also regions with timelike signatures (where $\Lie_n \tl < 0$). Such regions
appear to be associated with extreme conditions where new horizons form outside
of old ones. Though these so-called timelike membranes probably cannot be viewed
as horizons (they are traversible in both directions), they can evolve directly
into/from DHs and so need to be kept in mind during the following discussions
as possible quasilocal boundaries.  That said, MTTs  will usually be excluded
from consideration in the following as they introduce complications such as
non-trivial intersections with any foliation $\Sigma_t$ of the spacetime (see
\cite{bbgv} for many examples of these complications). 

\subsection{Horizon Geometry}
\label{horizons-quant}

In the preceeding section, we have discussed many different local horizon
formulations.  Here, we would like to examine the geometrical properties of
the horizons, focussing in particular on those which are common to all the
formulations.  Initially, we will restrict our attention to two-surfaces $S$
of zero expansion which foliate a three-surface $H$.  The intrinsic geometry
of the horizon is completely determined by the two-metric $\tq_{ab}$ and the
corresponding area two-form $\vS$ introduced previously.  The extrinsic
geometry of the surface is partially characterized by the extrinsic curvatures
($k_{ab}$) of $S$ along the null directions,
\begin{eqnarray}
  \frac{1}{2} \theta_{(\ell)} \tq_{ab} + \tq_{ac} \tq_{bd} \sigma_{(\ell)}^{cd}
  &=& k_{(\ell) ab} := \tq_{a}^{\; \; c} \tq_{b}^{\; \; d} 
 \nabla_{c} \ell_{d} \quad \mbox{and} \nonumber \\
 \frac{1}{2} \theta_{(n)} \tq_{ab} +  \tq_{ac} \tq_{bd} \sigma_{(n)}^{cd} 
 &=&  k_{(n) ab} :=  \tq_{a}^{\; \; c} \tq_{b}^{\; \; d} 
 \nabla_{c} n_{d}  \, .  \label{kex}
\end{eqnarray}
Here, $\tl$ and $\tn$ are the expansions of the null normals, introduced
previously, while $\sigma_{(\ell)}^{ab}$ and $\sigma_{(n)}^{ab}$ are the
shears along $\ell$ and $n$ respectively.  Since $\ell$ and $n$ are null
normals to the surface $S$ they are, by definition, twist free.  The
expansion of $\ell$ is required to vanish on all the horizons discussed above.
The other quantities will generically be non-zero except on an isolated
horizon when the shear $\sigma_{(\ell)}^{ab}$ must also vanish as there can be
no flux of gravitational energy across the horizon.  

The rest of the extrinsic geometry is described by the connection on the
normal bundle to $S$,
\begin{equation}\label{omega} 
  \omega_{a} = - n_c \nabla_a \ell^c \, .
\end{equation}
This quantity is closely associated in many of the formalisms with
physical characteristics of the horizons.  On an isolated
horizon, the surface gravity is obtained by contracting with $\ell$, whence
$\kappa_{(\ell)} = \ell^{a} \omega_{a} = - \ell^{a} n_c \nabla_a \ell^c$.
Additionally, the two-surface components of $\omega$, namely
\begin{equation}\label{tom}
  \tom_{a} = \tq_{a}^{\; \; b} \omega_{b} = 
- \tq_{a}^{\; \; b} n_c \nabla_{b} \ell^c \, , 
\end{equation}
are often associated with quasilocal angular momentum \cite{by1, ak, epp, eric,
ih}.  Specifically, on an axisymmetric isolated or slowly evolving horizon
with a symmetry generated by $\phi^{a}$, the angular momentum is given by
\begin{equation}\label{ang_mom}
  J_{IH}[\phi] = - \frac{1}{8 \pi G} \int_S \vS \phi^a \tom_a \, . 
\end{equation}

The definition of angular momentum on a dynamical horizon is slightly
different.  Since dynamical horizons are by definition spacelike, we can
introduce the timelike unit normal to the horizon $\hat{\tau}^{a}$ and the
unit spacelike normal to the leaves of the horizon, $\hat{r}^{a}$.  Then, the
angular momentum is constructed from these vector fields as
\begin{equation}\label{ang_mom_dh}
  J_{DH}[\phi] =  \frac{1}{8 \pi G} \int_S \vS \phi^a \hr^b \nabla_a \htau_b 
  \, .
\end{equation}
Then, it is straightforward to integrate the diffeomorphism
constraint over the horizon to obtain a flux law for angular momentum \cite{ak}.
Specifically, given a region of the horizon $\Delta H$ bounded by two
surfaces $S_2$ and $S_1$, the angular momentum satisfies
\begin{equation}\label{angflux}
  J_{2}[\phi] - J_{1}[\phi] = \int_{\Delta H} \mspace{-20mu} 
  \vH \left\{ T_{ab} \htau^a \phi^b 
+ \frac{1}{16 \pi G} P^{ij}_{\htau} \Lie_{\phi} q_{ij} \right\} \, .  
\end{equation}
In the above $\vH$ is the natural volume form on the horizon.  The two terms in
(\ref{angflux}) can then be interpreted as the flux of matter and gravitation
angular momentum through the horizon.

%%%%%%%%%%%%%%%%%%%%%%%%%%%%%%%%
% THE ACTION
%%%%%%%%%%%%%%%%%%%%%%%%%%%%%%%%
\section{The Action}
\label{actSect}

In this section we construct an action formulation for general relativity over
a quasilocal region of spacetime that includes a trapping horizon $H$ as one
of its boundaries.  To do this, we first introduce the spacetime $M$ of
interest, and then we specify appropriate boundary conditions on each boundary
of the spacetime.  Those on the horizon will be specified in terms of the
geometrical properties discussed in Section \ref{horizons-quant} while those
on the other boundaries are standard and well-known. Combining these with an
action whose bulk term is the usual Einstein-Hilbert term, we show that the
first variation of the action vanishes in the required way if the Einstein
equations hold in the bulk.  Thus, an action formulation exists for spacetimes
which have a horizon as a boundary.  Finally, by performing a Legendre
transform, we derive a corresponding quasilocal Hamiltonian from this action.
In the usual way, the form of this functional then provides a definition of
the energy associated with the region and its assorted boundaries.

\subsection{The spacetime and its boundaries} 
\label{ActionSetup}

Let $M$ be a finite region of spacetime with metric $g_{ab}$,
metric-compatible derivative $\nabla_a$, and four-volume form $\epsilon_{abcd}$
(also written as $\vM$).  We require $M$ to be time orientable and
diffeomorphic
to $[ 0,1 ] \times [ 0,1 ]  \times S^{2}$, where the four boundaries are:
\begin{enumerate}

\item a spacelike past boundary $\Sigma_{1}$ which represents the initial
configuration, 

\item a spacelike future boundary $\Sigma_{2}$ which represents the
final configuration,

\item a timelike outer boundary $B$, and  

\item an inner horizon boundary $H$ which is foliated by two-spheres
\footnote{
Strictly speaking the only property of $S^2$ that we will need in the following calculations is that
it is closed. That said it can be shown \cite{hayward} that in most physically
interesting situations, the Einstein equations along with energy conditions actually 
restrict the $S_v$ to being topologically $S^2$. Thus, for definiteness, we make this assumption.}
$S_v$
of zero outward null expansion ($\tl = 0$).  
\end{enumerate}	
\begin{figure}
\input{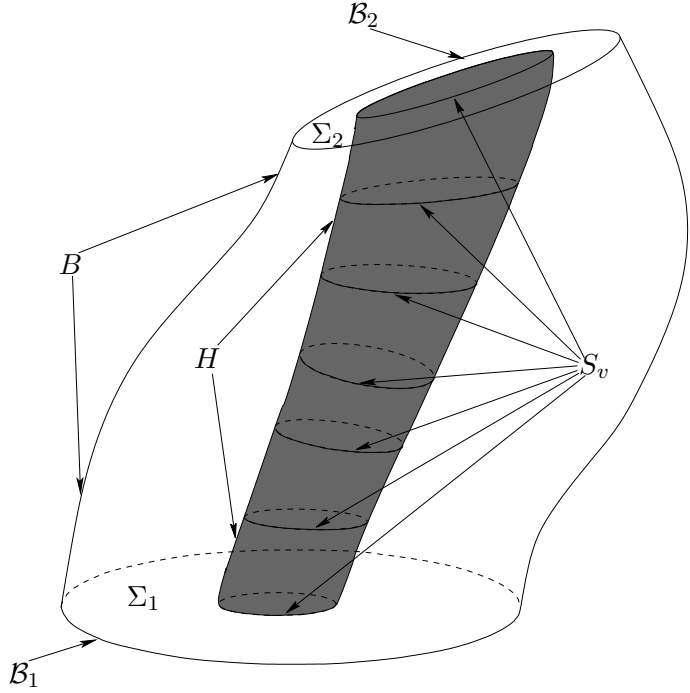}
\caption{A three-dimensional representation of $M$ along with its boundaries.}
\label{Fig1}
\end{figure}
The two-dimensional intersections of these boundaries are $\mathcal{B}_{1} =
\Sigma_{1} \cap B$,  $\mathcal{B}_{2} = \Sigma_{2} \cap B$, $S_{1} =
\Sigma_{1} \cap H$, and $S_{2} = \Sigma_{2} \cap H$. Figure \ref{Fig1}
provides a three-dimensional representation of the manifold $M$ and these
assorted boundaries.

In describing the precise boundary conditions on each of the surfaces and also
in the ensuing calculations, it will often be important to distinguish between
four-dimensional quantities and those defined exclusively in the various two-
and three-dimensional surfaces.  To this end, we introduce some notational
conventions.  Tensor fields which are elements of the tensor bundles of the
full spacetime will be labelled with lower case early alphabet Latin indices
$(a,b,c, \mbox{etc.})$. Tensors in the bundles associated with
three-dimensional hypersurfaces will be labelled with lower case mid-alphabet
Latin indices $(i, j, k, \mbox{etc.})$  and tensors in the bundles associated
with two-dimensional surfaces will be labelled with upper case Latin indices
$(A, B, C, \mbox{etc.})$.  Additionally, we will need to map tensors between
the tensor bundles associated with the different manifolds.  We will write all
pull-back/push-forward operators as $p$ with indices added to indicate which
spaces it maps between. Thus, the four-dimensional metric $g_{ab}$ pulled back
into a two-dimensional surface would be written $p_{A}^{\; \; a}
p_{B}^{\;\; b} g_{ab}$, while a three-dimensional vector $v^i$ would
push forward to $p_{i}^{\;\; a} v^i$ in the full spacetime. Of course, as
indicated by the indices, only fully covariant tensors may be pulled back and
only fully contravariant tensors may be pushed forward in a metric independent
way. If we wish to project tensors of a different rank we must make use of the
metric, rendering the results metric dependent.
 
\begin{figure}
\input{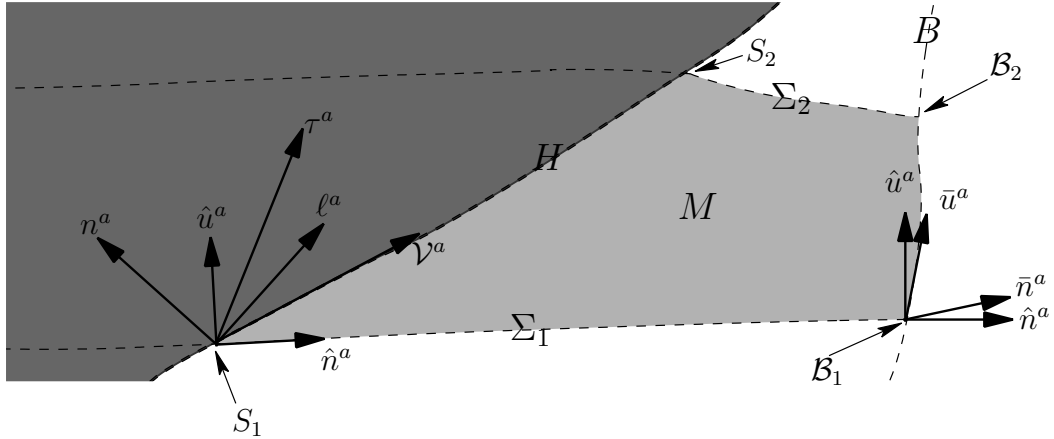}
  \caption{A two-dimensional representation of the spacetime containing $M$
  (the lightly shaded region), showing the past and future three-boundaries
  $\Sigma_1$ and $\Sigma_2$ along with the inner and outer three-boundaries
  $H$ and $B$. The intersections of these surfaces $S_{1,2}$ and
  $\mathcal{B}_{1,2}$ are also shown along with the orientations of the
  assorted normal vector fields for the case where $H$ is spacelike and
  expanding.  The darkly shaded region is the interior of $H$ -- the ``hole".
  Each point of this schematic is a two-sphere in the real $M$. }
\label{Fig2}
\end{figure}

With the preliminaries out of the way, we now describe in more detail the
various boundaries of $M$.  Figure \ref{Fig2} provides a two-dimensional
representation of the spacetime $M$ and its boundaries.  Surface normals
and other important vector fields are also shown.

\subsubsection*{Initial and final boundaries : $\Sigma_{1}, \Sigma_{2}$}

\noindent Denote the future directed unit normal vectors to the spacelike
boundaries $\Sigma_{1,2}$ by $\hat{u}^a$. Then, the three-metric on the
surfaces $\Sigma$ is given by $h_{ij}$ which satisfies 
$h^{ab} := p_{i}^{\;\; a} p_{j}^{\;\; b} h^{ij} = g^{ab} + \hu^a \hu^b$. 
The metric compatible derivative operator is
$D_{i}$. We obtain the three-volume form by contracting the four-volume form with
the normal $\hat{u}^a$.%
\footnote{The volume form is induced on every submanifold by Stokes theorem.
It can be found by contracting $\epsilon_{abcd}$ with the outward normal on
the \textit{last} index.}  %
Thus, $\epSigma_{abc} = \epsilon_{abcd} \hat{u}^d$ from which it follows that
$\vM = \mathbf{\hat{u}} \wedge \! \! \vSigma$.  Finally, the extrinsic
curvature of $\Sigma$ in $M$ is given by $K_{(\hat{u})}^{ab} = h^{ac} h^{bd}
\nabla_{c} \hat{u}_d$.%
\footnote{For those familiar with the work of Brown and York \cite{by1}, note
that they define extrinsic curvature with the opposite sign.}

\subsubsection*{Outer boundaries : $B$, $\mathcal{B}_{1}, \mathcal{B}_{2}$}

\noindent In keeping with the standard quasilocal action analyses, we take the
outer boundary $B$ to be timelike with $\bar{n}^{a}$ as its outward pointing
spacelike unit normal.  Then, the three-metric on $B$ is  $\gamma_{ij}$, where
$\gamma^{ab} =  p_{i}^{\;\; a} p_{j}^{\;\; b} \gamma^{ij} = g^{ab} - \bar{n}^a
\bar{n}^b$ and it defines the metric compatible derivative $\Delta_i$. The
extrinsic curvature $K_{(\bar{n})}^{ab} = \gamma^{ac} \gamma^{bd} \nabla_{c}
\bar{n}_d$ and the volume form is taken to be $\vB_{abc} = -\epsilon_{abcd}
\bar{n}^d$, so that $\vM = \mathbf{\bar{n}} \wedge \!  \! \vB$. 
\footnote{Note that this is the opposite of the volume form induced by Stokes
theorem. Consequently we will introduce a minus sign when integrating an exact
derivative out to the boundary $B$.}

The outer boundary $B$ intersects $\Sigma_{1}$ and $\Sigma_{2}$ on the
spacelike two-surfaces $\mathcal{B}_{1}$ and $\mathcal{B}_{2}$.  We do not
require that this intersection be orthogonal (i.e. there is no requirement
that $\hat{u} \cdot \bar{n} = 0$).  Thus, we introduce two additional normals:
$\hat{n}^a$, the outward unit spacelike normal to $\mathcal{B}_{1,2}$ in
$\Sigma_{1,2}$, and $\bar{u}^{a}$, the future directed timelike unit normal to
$\mathcal{B}_{1,2}$ in $B$ (as illustrated in figure \ref{Fig2}). Defining
$\eta = -\hat{u} \cdot \bar{n}$, the unit normals are related by 

\begin{equation}\label{bubn} \bu^a = \sqrt{1+ \eta^2} \, \hu^a + \eta \, \hn^a
\quad \mbox{and} \quad \bn^a = \eta \, \hu^a  + \sqrt{1+ \eta^2} \, \hat{n}^a
\, .  \end{equation}
Clearly, when $\eta = 0$, $\bu^a = \hu^a$ and $\bn^a = \hn^a$ and so the
$\Sigma/B$ intersection is orthogonal. 

The two-metric on $\mathcal{B}_{1,2}$ is denoted $\bar{\gamma}_{AB}$ and its
inverse pushes forward to $\bar{\gamma}^{ab} = h^{ab} - \hat{n}^{a}
\hat{n}^{b} = \gamma^{ab} + \bar{u}^{a} \bar{u}^{b}$.  The derivative operator
compatible with this metric is $d_{A}$.  The volume form $\vBt$ on
$\mathcal{B}_{1,2}$, is given by $\vM = \mathbf{\bar{u}} \wedge
\mathbf{\bar{n}} \wedge \vBt = \mathbf{\hat{u}} \wedge \mathbf{\hat{n}}
\wedge\vBt$.%

\subsubsection*{Inner boundaries: $H$, $S_{1}$, $S_{2}$}

\noindent A discussion of the inner boundary $H$ is more involved than that of
the outer boundary. Since this boundary is required to be a trapping horizon:
\begin{enumerate}

\item $H$ has additional structure in the form of a foliation into the
$\tl=0$ spacelike two-surfaces and 

\item there is no restriction on the signature of $H$ and so it can be
spacelike, null, timelike, or conceivably some mixture of the three.

\end{enumerate}
We label the foliation two-surfaces $S_{v}$, where $v$ is a foliation
parameter.  Furthermore, we assume that $S_1 \equiv \Sigma_1 \cap H$ and $S_2
\equiv \Sigma_2 \cap H$ are the initial and final leaves of that foliation.
That is, $S_1 \equiv S_{v_1}$ and  $S_2 \equiv S_{v_2}$ and $v_{1} \leq v \leq
v_{2}$. The foliation parameter naturally induces the one-form $dv \in T^{\ast} \! H$ normal to
the horizon 2-surfaces. In addition, we introduce a vector field
$\frac{\partial}{\partial v}$ tangent to the horizon and satisfying
$\Lie_{\left( \partial/\partial v \right)} v =1$.  It is not unique, but will be 
useful in describing how the cross-sections $S_{v}$ evolve.

Since the horizon is foliated, we can introduce normals to the two-surfaces.
It is most convenient to work with the null normals $\ell_a$ and $n_a$ to the
surfaces $S_{v}$.  This will allow us to treat timelike, spacelike and null
horizons $H$ together and also make it easier to impose the condition that
$\tl = 0$.  As shown in figure \ref{Fig2}, we let $\ell^a$ be outward (towards
$B$) and future pointing while $n^a$ will be inward (away from $B$) and also
future pointing.  Furthermore, we require the standard normalization $\ell
\cdot n = -1$.  This, however, still leaves one rescaling degree of freedom in
the null vectors which will be important during the calculations.
Irrespective of how this is chosen, the two-metric $\tq_{AB}$ satisfies
$\tq^{ab} = p_{A}^{\;\; a} p_{B}^{\;\; b} \tq^{AB}= g^{ab} + \ell^a n^b + n^a
\ell^b$. As on the other two-surfaces, the metric compatible derivative will
be denoted $d_A$.  Additionally, in Section \ref{horizons-quant} we have
introduced the extrinsic curvatures $k_{(\ell)}^{AB}$ and $k_{(n)}^{AB}$ of
the null vectors, as well as the one form $\tom_{A}$ which contains the
angular momentum information.

On the initial and final cross sections of the horizon, we can relate the null
normals $\ell$ and $n$ to those inherited from $\Sigma_{1}$ and $\Sigma_{2}$.
Specifically, we have the timelike normal $\hu^a$ inherited from the $\Sigma_1
/ \Sigma_2$ boundaries and the outward-pointing spacelike normal $\hn^a$ to
$S_1 / S_2$ in $\Sigma_1 / \Sigma_2$ as shown in figure \ref{Fig2}. Then
\begin{equation}\label{lnu} 
  \ell^a = \zeta ( \hu^a + \hn^a ) \hspace{1cm} \mbox{and} \hspace{1cm} 
	n^a = \frac{1}{2 \zeta} ( \hu^a - \hn^a ) \, ,
\end{equation}
for some function $\zeta$.  Furthermore, we can also write the two-metric as
$\tq^{ab} = g^{ab} + \hu^a \hu^b - \hn^a \hn^b$.

We now turn to characterizing the three-dimensional horizon $H$.  We begin by
introducing a normal to $H$ in $M$, which we denote $\tau_a$,  and a vector
field $\cV^a$ that is both tangent to $H$ (so that $\cV^a \tau_a = 0$) and
perpendicular to the $S_v$.  Both are assumed to be future pointing (in the sense that 
$\cV \cdot n$ and $\tau \cdot n$ are strictly negative).  Since
the horizon is not of fixed signature, there is no natural normalization for
these structures,  however, it will often be convenient to normalize both
relative to the foliation label $v$.  Thus, we require
\begin{equation}\label{vrelation}
  \Lie_{\mathcal{V}} v = 1 \, \Leftrightarrow \frac{\partial}{\partial v} =
	\mathcal{V} + \tilde{V} \quad \mathrm{where} \quad \tilde{V} \in TS_{v} \, .
\end{equation}
and furthermore, 
\begin{equation}\label{vtaunorm}
  \tau \cdot \tau = - \mathcal{V} \cdot \mathcal{V} \, .
\end{equation}
For those familiar with dynamical horizons \cite{ak}, when $H$ is spacelike,
$\mathcal{V}^a$ is spacelike and proportional to the unit vector $\hr^a$ of
those papers.  Similarly $\tau_a$ becomes timelike and is proportional to the
unit $\htau_a$ of those papers. By contrast, when $H$ is null, the fields are
degenerate and are both parallel to $\ell$.

For a given two-surface $S_{v}$, we have introduced two sets of
(future-pointing) normal vector fields: $(\ell^{a}, n^a)$ and $(\mathcal{V}^a
, \tau^a)$.  Clearly these must be related.  To this end, we introduce a
scalar field $C$ on $H$ chosen so that $\ell_{a} + C n_{a}$ is normal to $H$,
and hence proportional to $\tau_{a}$.  It follows that $\ell^{a} - C n^{a}$
will then lie in $H$ and be proportional to $\mathcal{V}^a$.  Therefore, we
can write
\begin{equation}\label{alphaDef} 
  \mathcal{V}^a = \alpha (\ell^a - C n^a ) \mspace{18mu} \mbox{and} 
	\mspace{18mu} \tau_a = \alpha (\ell_a + C n_a) \, , 
\end{equation}
for some $\alpha > 0$ on $H$.
The sign of $C$ naturally relates to the signature of
the induced metric $q_{ij} = p_{i}^{\; \; a} p_{j}^{\; \; b} g_{ab}$ on $H$;
specifically, if $C<0$, $C=0$, or $C>0$, $H$ is respectively timelike, null,
or spacelike.  Finally, the orientation of the volume form on the horizon is chosen so that
\begin{equation}\label{horizonvol}
  \vM = \bta \wedge \bdv \wedge \vS =  - \bfl \wedge \bfn \wedge \vS \, . 
\end{equation}

Note that we have said very little about the geometry of $H$ itself.  The
reason for this is that the ensuing calculations will largely focus on
two-surface rather than three-dimensional quantities. This is the most
convenient way to proceed given that $H$ does not have a fixed signature and
so may change between spacelike, null, and timelike signatures.  Similarly, in
order to avoid having to deal with the timelike, null, and spacelike
quantities seperately we choose not to work with normalized quantities such as
$\htau_a$ and $\hr^a$ which are ill-defined if $H$ becomes null.  

\subsection{Boundary conditions}
\label{sbh}

In order to obtain a well-defined action principle, it is necessary to specify
boundary conditions for all of the boundary manifolds presented in the
previous section.  First, 
on the initial and final boundaries $\Sigma_{1}$ and $\Sigma_{2}$, as well as
the outer boundary $B$, we fix the three metric.  Similarly on the
two-dimensional intersections $\mathcal{B}$, we fix the two-metric.  Thus, we
require that 
\begin{equation}\label{metricfix}
  \delta h_{ij} = 0 \, , \quad \delta \gamma_{ij} = 0 \, \quad \mathrm{and} 
	\quad \delta \bar{\gamma}_{AB} = 0 \, .
\end{equation}
These are by no means the only permissible boundary conditions, but they are
by far the most common.  See \cite{by2} for a discussion of alternative
boundary conditions.

The situation on $H$ is more complicated and consequently more interesting. We
start with a defining condition, $\tl = 0$, and so expect it to play an
important role in any calculations --- otherwise the resulting action
formulation will in no way be tied to horizons. With this principle in mind we
reject the idea of fixing the full three-metric $q_{ij}$ as this would give a
well-defined action principle regardless of the value of $\tl$.
Geometrically, it would also seem to be overly profligate compared to what is
done on the other boundaries. For example, for $\Sigma_1$, the geometry is
fully specified by its intrinsic metric, $h_{ij}$, and extrinsic curvature,
$K_{(\hu)}^{ij}$.  Each of these has six components, so for $\Sigma_1$ with
standard boundary conditions, six out of a possible twelve geometric
components are fixed.  Now, if we fix $q_{ij}$ on $H$ as well as $\tl=0$ we
would essentially have fixed seven of those components.  By contrast, the path
that we follow in this paper will fix $\tl = 0$ along with just five other
components of the surface geometry on $H$, and each of those conditions will
play a crucial role in the calculations. 

$H$ is also different from the other boundaries in that it comes with a
foliation. This foliation is not just an optional structure imposed on $H$,
but is crucial to its very definition.  Without the foliation, the key $\tl =
0$ condition is not well defined and, furthermore, for a given (spacelike) $H$
this foliation is unique \cite{gregabhay}.  Given this intricate association
of $H$ and its foliation, it is perhaps not surprising that it also features
prominently in our boundary conditions on $H$. We fix:
%

%
%As such we may decompose its geometry with respect to the $S_v$. 

%To start, note that the foliation surfaces $S_v$ play an important role in
%specifying a horizon (unlike the situation on $B$ where there is no
%foliation). The two-surfaces are used in the construction of the null normal
%$\ell_a$ which is required to be expansion free.  Furthermore, we have access
%to the two-metric $\tq_{AB}$ and angular momentum one-form $\tom_A$ defined on
%those surfaces.  These quantities on $S_{v}$ have a clear physical
%interpretation: they describe the area and angular momentum of the two
%dimensional cross sections of the horizon.  This suggests that we fix:
%
\begin{enumerate} 

\item $\theta_{(\ell)}= 0$, 

\item the spacelike two-metric $\tq_{AB}$ on each $S_v$
(which also fixes $\vS$), 

\item $\tom_A = - p_A^{\; \; a} \, n_b \nabla_a \ell^b$ (the
connection on the normal bundle to each $S_v$), and

\item $\underleftarrow{\ell}_i = p_{i}^{\; \; a} \ell_a$, the pull-back of
$\ell_a$ to $T^\ast \mspace{-2mu} H$.

\end{enumerate}
Then in the action formulation, the first variation of each of these
quantities is zero. The foliation surfaces $S_v$ as well as their coordinate
label $v$ are also kept fixed under these variations.  

The first condition is simply one of the defining conditions for a trapping or
dynamical horizon. Note however, that other defining conditions such as
$\theta_{(n)} < 0$, $\Lie_n \theta_{(\ell)} < 0$, or  $H$ being spacelike, are
not imposed. Thus, these boundary conditions apply equally well to dynamical
horizons or any of the varieties of trapping horizons,  including those
usually taken to represent white holes or inner black hole horizons
\cite{hayward}. It also includes three-dimensional surfaces traced by smoothly
evolving apparent horizons.%
\footnote{However,  we do not allow for situations where
$\Lie_n \tl$ changes sign and so $C$ diverges, such as the spacetimes
considered in \cite{bbgv}. These situations can be dealt with fairly easily 
by sectioning the horizon into spacelike/timelike sections over which $C$ is well-defined,
but for simplicity we will not consider these cases here.}

The second condition fixes the intrinsic geometry of each of the $S_v$
two-surfaces and so is somewhat similar to the usual fixing of the boundary
metric.  The difference, of course, is that here only the two-geometries are
fixed.  The rest of the three-geometry which gives the details how the
two-surfaces evolve into each other, including the signature of $H$, is left
free. Thus, while we know quantities such as the curvature scalar and area of
the two-surfaces, the boundary conditions won't fix such things as the shear
of $H$, or indeed any other locally defined rate of change. 

The third condition fixes another part of the extrinsic geometry of $S_v$ by
specifying the connection on the normal bundle.  Equivalently, it may be
thought of as fixing the angular momentum of the horizon.  Combined with the
previous conditions, we see that the boundary conditions on $H$ are mixed in
the sense that they fix parts of both the intrinsic and extrinsic (two)
geometry of the $S_v$. This contrasts with the conditions on $B$ and
$\Sigma_{1,2}$ which exclusively fixed the intrinsic (three) geometry.  It is
interesting to note that Epp \cite{epp} presented counting arguments which
suggest that knowledge of the intrinsic geometry of a two-surface plus the
curvature on its normal bundle should be sufficient to uniquely determine its
embedding in a four-dimensional space of constant curvature.  Thus, in this
sense, the information that we have specified on $S_v$ can be thought of as a
complete set of data that fully specifies its geometry.

Finally, we come to the fourth term. At first glance it may appear that we are
fixing a seventh ``component" of the geometry (given that $\tl=0$ is one,
$\tq_{AB}$ gives three, and $\tom_A$ gives two). However, that is not the
case.  Recall that when defining $\ell_a$ we left its normalization (and so
that of $n_a$) unspecified.  This condition removes that degree of freedom.
Hence, it is essentially equivalent to a requirement like the one that
specifies that extrinsic curvatures should be defined using unit-normal
vectors to surfaces.  Just as such a definition is necessary so that a
discussion of extrinsic curvature is meaningful, here we need a fix to give
$\tom_A$ a geometric meaning --- otherwise it would change with rescalings
of $\ell_a$.  Keep in mind that we only fix the pullback of $\ell$ to $H$, not
$\ell_a$ itself (which would have geometric significance).

\subsection{The quasilocal action and its variation}
\label{act}

Having defined $M$ and established this notation, we turn to the action
itself.  The action for the region $M$ and its boundaries is given by
\begin{eqnarray}\label{IH}
  I_H & = &  \frac{1}{16 \pi G} \int_M \mspace{-10mu} \vM  \mathcal{R} 
  +  \frac{1}{8 \pi G} \int_{\Sigma_2-\Sigma_1} \mspace{-36mu} \vSigma K_{\hu} 
	\\
  & & -  \frac{1}{8 \pi G} \int_B \mspace{-6mu} \vB K_{\bar{n}} 
	+ \frac{1}{8 \pi G} \int_{\mathcal{B}_2-\mathcal{B}_1} 
	\mspace{-36mu} \vBt \sinh^{-1} (\eta) \nn \\
  & & + \frac{1}{8 \pi G} \int_H \mspace{-6mu}  
	\vS \wedge dv \left( \kappa_{\mathcal{V}} + \tilde{V}^A
	\tilde{\omega}_A \right) \,   
  - \frac{1}{8 \pi G} \int_{S_2-S_1} \mspace{-32mu} \vS 
	\ln \zeta \, , \nn
\end{eqnarray}
where $\mathcal{R}$ is the Ricci scalar over $M$, $\kappa_{\mathcal{V}}$ is
given by
\begin{equation}
  \kappa_{\mathcal{V}} = \mathcal{V}^{a} \omega_{a} = - \mathcal{V}^{a} n_{b}
  \nabla_{a} \ell^{b} \, , 
\end{equation}
and $\tilde{V}^{a}$ is defined in (\ref{vrelation}). Equivalently, $
\kappa_{\mathcal{V}} + \tilde{V}^A \tilde{\omega}_A =
\left(\frac{\partial}{\partial v} \right)^{i} \omega_{i}$, and so we see that
this action explicitly depends on both the foliation and the associated vector
field $\left(\frac{\partial}{\partial v} \right)$ (though in the upcoming
calculations, we will see that this is essentially a gauge dependence only as
the variation of $I_H$ will vanish whatever the value of these coordinate
quantities).

The bulk, $\Sigma$, $B$, and $\mathcal{B}$ boundary terms are standard and
found in all discussions of the gravitational action. The integral of the
Ricci scalar over $M$ is the Einstein-Hilbert bulk term while the integrals
over $\Sigma_{1}$, $\Sigma_{2}$, and $B$  are the usual extrinsic curvature
terms which must be included when $M$ has timelike/spacelike boundaries (see,
for example, \cite{wald}). The fourth term may be a little less familiar, but
is also well known \cite{ghayward}. This ``corner" term is necessary when the
$\Sigma$ boundaries do not orthogonally intersect $B$; that is $\hu^{a}
\bn_{a} \neq 0$.  The final two terms arise at the horizon boundary and so are
new, though we immediately note that the last is a ``corner'' term between $H$
and $\Sigma_{1,2}$.

Evaluating the first variation of this metric is a non-trivial calculation.
Most readers will not be interested in the details and so
they are relegated to Appendix \ref{app1}. Here, we just quote the final
result: 
\begin{eqnarray} 
  \delta I_H &=& 
	\frac{1}{16 \pi G} \int_M \mspace{-10mu} \vM G_{ab} \delta g^{ab} 
	- \frac{1}{16 \pi G} \int_{\Sigma_2-\Sigma_1} \mspace{-36mu} \vSigma 
	\left( K_{\hu}^{ij} - K_{\hu} h^{ij} \right) \delta h_{ij}  \label{dIH} \\ 
	& & + \frac{1}{16 \pi G} \int_B \mspace{-6mu} \vB 
	\left( K_{\bar{n}}^{ij} - K_{\bar{n}} \gamma^{ij} \right) \delta \gamma_{ij} 
	+ \frac{1}{8 \pi G} \int_{\mathcal{B}_2-\mathcal{B}_1} \mspace{-32mu}
		 (\delta \vBt) \sinh^{-1} (\eta) \nonumber \\ 
	& & + \frac{1}{16 \pi G} \int_H \mspace{-6mu} \bdv \wedge \vS  
	\left(\tilde{s}^{AB} \delta \tq_{AB} \right) 
	+ \frac{1}{8 \pi G} \int_{S_2-S_1} \mspace{-32mu} \delta \vS 
	\left( 1 - \ln \zeta \right) \nn \\ 
	& &  + \frac{1}{8 \pi G} \int_H  \mspace{-6mu} \left\{ \vS \wedge \left(
	(\delta \lpb) \theta_{(n)} - (\delta \npb) \theta_{(\ell)} \right) 
	+ \tilde{V}^A \,\delta \left(\bdv \wedge \vS \, \tilde{\omega}_{A} \right) 
	- \delta \left( 2 \tl \npb \wedge \vS \right) \right\} \nn \, .  
\end{eqnarray}
Here, the quantity $\tilde{s}^{AB}$ is given by
\begin{equation}\label{tildes}
  \tilde{s}^{AB} = ( k_{\tau} +  \kappa_{\mathcal{V}}) \tq^{AB} 
  - k_{\tau}^{AB} \, ,
\end{equation}
while $\lpb$ and $\npb$ respectively denote the pull-backs of $\ell_a$ and $n_a$
into $H$.  

As for the action itself, the boundary terms on $\Sigma_{1 \, , \, 2}$ and
$\mathcal{B}_{1 \, , \, 2}$ are completely standard and it is clear that
fixing $\delta h_{ij} = 0 $ and $\delta \gamma_{ij}= 0$ will ensure that these
vanish. Similarly, it is not hard to see that our boundary conditions from
Section \ref{sbh} are sufficient to ensure the terms on $H$ and $S_{1 \, , \,
2}$ vanish. Thus, $\delta I_H = 0$ if and only if Einstein's equations hold in
the bulk and so this action is well-formulated. 

Finally we note that, as in most Lagrangian analyses, the action that we have
considered is not the only action compatible with the given boundary
conditions. In fact, any functional of the form
\begin{equation}\label{freedom}
  I'_H = I_H + \left( \mbox{functional of the fixed data} \right) \, , 
\end{equation}
will work equally well, since the variation of a functional of the fixed data
is zero. The implications of this freedom will be discussed in more detail in
Section \ref{DynHam}.

\subsection{Weak boundary conditions}
\label{wbh}

In the previous section, we have seen that the action principle is
well-formulated for a manifold with a trapping horizon boundary.  In the
calculation, the boundary conditions have been imposed in the standard way;
namely by requiring that variations of fixed quantities vanish.  However for
boundary conditions such as those on $B$, it has recently been shown
\cite{BLY,bf1} that a weaker form of the boundary conditions can be used
instead.  These weak boundary conditions only fix the data up to
diffeomorphisms.  The motivation for this is the principle that  physical
theories should be generally covariant.  In the presence of boundaries, we
take this to mean that while we may (partially) specify the geometry of those
surfaces, we should not tie that geometry to specific points of $\Sigma_{1 \,
, \, 2}$, $\mathcal{B}_{1 \, , \, 2}$, $S_{1 \, , \, 2}$, $B$ or $H$ --- just
as in general relativity spacetime geometry isn't tied to points of an
underlying manifold (or equivalently a coordinate system). 

We now show that the boundary conditions of Section \ref{sbh} can
also be applied in this weak form ---  up to diffeomorphisms which preserve
the underlying manifold structure of the boundaries.  Therefore, we require
that
\begin{equation}
  \delta h_{ij} = \Lie_{Y} h_{ij} \, , \quad 
	\delta \gamma_{ij} = \Lie_{Z} \gamma_{ij}
	\quad \mathrm{and} \quad 
	\delta \bar{\gamma}_{AB} = \Lie_{\bar{y}} \bar{\gamma}_{AB}
\end{equation}
where $Y^i \in T \Sigma_{1,2}$, $Z^{i} \in TB$ and $\bar{y}^{A} \in T
\mathcal{B}_{1,2}$.  Since the boundaries $\mathcal{B}_{1,2}$
must be preserved (and so mapped into themselves), we impose the further 
restrictions that
\begin{equation}
  Y^{a} = Z^{a} = \bar{y}^{a} \;\;\mathrm{on}\;\; \mathcal{B}_{1,2}\, ,
\end{equation}
where each of these is the push-forward of the appropriate lower-dimensional
vector field.  

Similarly, at the horizon, we allow diffeomorphisms which preserve the
geometric structure, which in this case includes the foliation $S_{v}$. Thus,
such maps should send all points in a given leaf $S_{v}$ into a second leaf
$S_{\phi_X(v)}$ for some function $\phi_X(v) : [v_1, v_2] \rightarrow [v_1 ,
v_2]$.  The most general vector fields generating a one-parameter family of
diffeomorphisms of this type take the form
\begin{equation}
  X^{i} = x_{o} \mathcal{V}^i + p_{A}^{\;\; i} \tilde{x}^A 
  \;\;\mathrm{where}\;\; 
	\tilde{x}^A \in T S_{v} \, , \label{X1}
\end{equation}
where $x_o = x_o(v)$ is a function of the foliation parameter $v$, and

\begin{equation}
  x_{o} = 0 \;\;\mathrm{and}\;\; Y^{i} = \tilde{x}^{i} \;\;\mathrm{on}\;\; 
	S_{1,2} \, .
\end{equation}
Then, the weak form of the boundary conditions fixes :
\begin{equation}
  \delta \tilde{q}_{AB} = \Lie_{X} \tilde{q}_{AB} \, , \quad
	\delta \tilde{\omega}_{A} = \Lie_{X} \tilde{\omega}_{A} \, , \quad
	\delta \lpb_{i} = \Lie_{X} \lpb_{i} \, \quad \mathrm{and} \quad
	\theta_{(\ell)} = \delta \theta_{(\ell)} = 0 \, .
\end{equation}

The action principle will remain well-defined if we also enforce the
diffeomorphism constraint on each of the boundary three-surfaces.  This is not
a big assumption as these are naturally induced if the Einstein equations hold
in the bulk.  Given a normal $\chi_a$ and a tangent vector $X^a$ to a surface,
the (vacuum) diffeomorphism constraint on that surface is $X^a G_a^{\; \; b}
\chi_b = 0$.  Then on $\Sigma_{1 \, , \, 2}$ and $B$ this constraint says that
\begin{equation}\label{diffeoCon}
  D_{i} \left( K_{\hat{u}}^{ij} - K_{\hat{u}} h^{ij} \right) = 0 
	\quad \mathrm{and} \quad 
	\Delta_{i} \left( K_{\bar{n}}^{ij} - K_{\bar{n}} \gamma^{ij} \right) = 0 \,
	,
\end{equation}
where we respectively take $\hu$ and $\hn$ as the normal vectors. 

The situation on $H$ is a little more complicated since its uncertain
signature means that we cannot utilize unit normals or assume the existence of
a (unique) inverse metric.  The most useful form for our purposes is found by
taking $\tau_a$ (defined in equation (\ref{alphaDef})) as the normal and
evaluating 
\begin{equation} 
  X^a G_{a}^{\; \; b}\tau_b = 0 \, ,
\end{equation}
where we continue to assume that $T_{ab} = 0$. 
For the current calculation we further specialize this by integrating over
$S_v$. Then, as noted in \cite{prl}:
\begin{equation}
  \frac{1}{8 \pi G} \int_{S_v} 
	\left\{ \kappa_{\mathcal{V}} \Lie_X \vS
	+ \tilde{x}^A \Lie_{\mathcal{V}} [ \vS \tom_A ] \right\}  
	= \frac{1}{16 \pi G} \int_{S_v} \vS  \left\{ k_{\tau}^{AB} \Lie_X \tq_{AB} 
	+ 2 \alpha C \Lie_X \theta_{(n)} \right\} \, .  \label{prlCon} 
\end{equation}
A full derivation of this result will be found in \cite{bfbig}, but for now a
few lines of algebra puts it into a more useful form for the current
calculation:
\begin{equation}\label{horCon}
  \frac{1}{8 \pi G} \int_{S_2 - S_1} \mspace{-32mu} \vS (\tx{^A} \tom_A) 
	= \frac{1}{8 \pi G} \int_H \mspace{-6mu} \left\{ \bdv \wedge 
  \vS \left( \frac{\tilde{s}^{AB}}{2} \Lie_X \tilde{q}_{AB} \right)
	- \vS \wedge (\Lie_{X} \lpb) \theta_{(n)} - \tilde{V}^{A} \Lie_{X}
	( \bdv \wedge \vS \tom_{A} ) \right\} \, .  
\end{equation}

With the weaker boundary conditions, the variation of the action (\ref{dIH})
becomes
\begin{eqnarray} 
  \delta I_H &=& 
	\frac{1}{16 \pi G} \int_M \mspace{-10mu} \vM G_{ab} \delta g^{ab} 
	- \frac{1}{16 \pi G} \int_{\Sigma_2-\Sigma_1} \mspace{-36mu} \vSigma 
	\left( K_{\hu}^{ij} - K_{\hu} h^{ij} \right) \Lie_{Y} h_{ij}  \\ 
	& & + \frac{1}{16 \pi G} \int_B \mspace{-6mu} \vB 
	\left( K_{\bar{n}}^{ij} - K_{\bar{n}} \gamma^{ij} \right) 
	\Lie_{Z} \gamma_{ij} 
	+ \frac{1}{8 \pi G} \int_{\mathcal{B}_2-\mathcal{B}_1} \mspace{-32mu}
		 (\Lie_{y} \vBt) \sinh^{-1} (\eta) \nonumber \\ 
	& & + \frac{1}{16 \pi G} \int_H \mspace{-6mu} \bdv \wedge \vS  
	\left(\tilde{s}^{AB} \Lie_{X} \tq_{AB} \right) 
	+ \frac{1}{8 \pi G} \int_{S_2-S_1} \mspace{-32mu} \Lie_{\tilde{x}} \vS 
	\left( 1 - \ln \zeta \right) \nn \\ 
	& &  + \frac{1}{8 \pi G} \int_H  \mspace{-6mu} \left\{ \vS \wedge \left(
	(\Lie_{X} \lpb) \theta_{(n)} - (\delta \npb) \theta_{(\ell)} \right) 
	+ \tilde{V}^A \,\Lie_{X} \left(\bdv \wedge \vS \tilde{\omega}_{A} \right) 
	- \delta \left( 2 \npb \theta_{(\ell)} \wedge \vS \right) \right\} \nn \, .  
\end{eqnarray}
The diffeomorphism constraints of $\Sigma_{1,2}$ and $B$ allow us to rewrite
the boundary terms on those surfaces as exact derivatives.  These can then be
integrated to give contributions on $\mathcal{B}_{1,2}$ and $S_{1,2}$.
Similarly, the horizon constraint (\ref{horCon}) can be used to replace the
terms on $H$ with a term on $S_{1,2}$.  Therefore, we arrive at
\begin{eqnarray}
   \delta I_H &=& 
	 \frac{1}{16 \pi G} \int_M \mspace{-10mu} \vM G_{ab} \delta g^{ab}
	 + \frac{1}{8 \pi G} \int_{\mathcal{B}_2-\mathcal{B}_1} \mspace{-32mu}
	 \left\{ \left( \Lie_{y} \vBt \right) \sinh^{-1} (\eta) 
	 - \vBt  y^{a} \left(\hat{n}^{b} \nabla_{a} \hat{u}_{b} 
	 + \bar{u}^{b} \nabla_{a} \bar{n}_{b} \right) \right\} \nonumber \\
	 && \quad	+ \frac{1}{8 \pi G} \int_{S_2-S_1} \mspace{-32mu} \left\{
	 \left( \Lie_{\tilde{x}} \vS \right) \left(1 - \ln \zeta \right) 
	 + \vS  \tilde{x}^{a} \left( \hat{n}^{b} \nabla_{a} \hat{u}_{b} 
	 + \bar{n}^{b} \nabla_{a} \bar{u}_{b} \right) \right\} 
\end{eqnarray}

Finally, we can make use of the relationship between $\eta$, $\hat{n}$,
$\hat{u}$ and $\bar{n}$, $\bar{u}$ given in (\ref{bubn}) to show that the
terms on $\mathcal{B}_{1,2}$ evaluate to an exact derivative which then
integrates to zero.  Similarly the terms on $S_{1,2}$ are exact as can easily
be seen by using the relationship between $\zeta$, $\hat{n}$, $\hat{u}$ and
$\ell$, $n$ given in (\ref{lnu}).  The only remaining term is then the bulk
term which vanishes for solutions to Einstein's equations.  Thus, we have shown
that the action principle remains well formulated when we enforce weak
boundary conditions while requiring that the diffeomorphism constraints hold on
the boundaries of the spacetime.

With these weak boundary conditions, the freedom in the action is vastly
reduced as boundary terms like those of (\ref{freedom}) must now be
covariantly defined. That said, it is clear that the parameter space of
possible actions remains large and, for example, includes all actions of the
form
\begin{equation}
I'_H = I_H + \int_{\Sigma_2 - \Sigma_1} \mspace{-32mu} \mathbf{F}^\Sigma (h_{ij}) 
                   + \int_{B} \mathbf{F}^B (\gamma_{ij}) 
                    + \int_H \mathbf{F}^H (\tq_{AB}, \tom_A, \lpb, \tl, v) \, , 
\end{equation}
where the $\mathbf{F}$ are required to be dimensionally correct, but are
otherwise freely defined volume forms with the displayed functional
dependencies. A lot has been written on possible boundary terms on $B$ (see
for example \cite{by1, epp, intrinsic}) but for now we simply note that the
mathematical formalism allows these freedoms and that they can only be further
reduced by imposing extra restrictions on the system. A good example of such a
restriction is the requirement that the evaluated action not diverge in
asymptotically flat spacetimes as $B$ is taken to infinity. In this case extra
boundary terms on $B$ not only may, but in fact \textit{must} be added to the
action to remove such divergences.

In this paper, the external boundary $B$ is not the chief concern and so we  will
not consider these terms further. Instead we focus on the internal boundary
$H$ and so consider actions that are equivalent to $I_H$ up to terms of the
form 
\begin{equation}\label{Hterm} 
		\Delta I = - \int dv E [\tq, \tom, \lpb, \tl, v]  \, ,  
\end{equation} 
where $E[\tq, \tom,\lpb, \tl, v]$ is a free functional of the fixed data with
dimension length.  The requirement that the extra boundary term break up in this
way is natural given the primary role played by the foliation of $H$. 
As we will now see, such terms are also very convenient as they allow for a 
clean Legendre transform of the action.

\subsection{Legendre transform to obtain a Hamiltonian}
\label{DynHam}

\begin{figure} 
\input{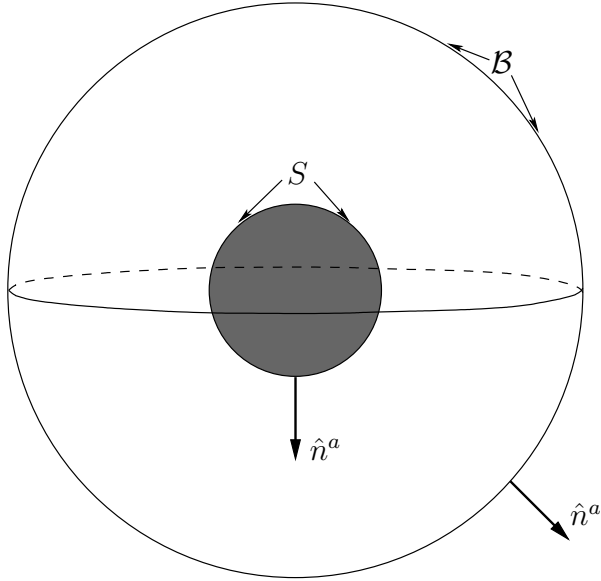} 
\caption{A three-dimensional volume $\Sigma_t$ along with its boundaries and
their normals.} 
\label{Fig0}
\end{figure}

In the previous sections it has been shown that the action $I_H$ (\ref{IH}) is
valid with either a strong or a weak fixing of the boundary conditions.  In
this section we introduce a notion of time to $M$ which then allows us to
perform a Legendre transform on the action and so derive a corresponding
Hamiltonian functional.

\subsubsection*{Flow of time and associated quantities}

\noindent Coordinate time must be externally imposed in general relativity. We
do this in the following manner. First, foliate $M$ with spacelike (partial
Cauchy) surfaces $\{ (\Sigma_t , h_{ij}) ; t_1 \leq  t \leq t_2 \}$;  these
will correspond to ``instants of simultaneity'' (see figure \ref{Fig0}).  For
simplicity, we assume that this foliation is compatible with all of the
structures that we have already introduced. Thus $\Sigma_{t_{1 \, , \, 2}} =
\Sigma_{1 \, , \, 2}$ and if $\{ \mathcal{B}_t ; t_1 \leq t \leq t_2 \}$ is
the induced foliation of $B$, then $\mathcal{B}_{t_{1 \, , \, 2}} =
\mathcal{B}_{1 \, , \, 2}$. On $H$ we require that the induced foliation be
compatible with $S_v$ and so it is natural to drop $v$ and henceforth use the
single parameter $t$ to characterize both foliations.   This extra structure
on $M$ and its boundaries allows us to extend the definitions of several
quantities that were previously only defined on initial and final boundaries
$\mathcal{B}_{1,2}$ and $S_{1,2}$. Specifically, with the help of $\Sigma_t$
we can define $\hu^a$ and $\hn^a$ everywhere on $B$ and $H$. Further $\bu^a$
and $\eta := - \hu \cdot \bn $ may be extended over $B$, while $\zeta$ may
be defined everywhere on $H$. 

In addition to a foliation parameter, we must also introduce a corresponding
notion of evolution.  This is provided by the evolution vector field $T^a$
which is assumed to be compatible with the foliation in the sense that 
\begin{equation}
  \Lie_T t = 1 \, . \label{Tt1}
\end{equation}
Thus, the flow generated by $T^a$ evolves leaves of the foliation into each
other in the required way. It is also assumed to be compatible with the
boundaries so it maps them into themselves; equivalently on $B / H$, $T^a \in
TB / H$ respectively.  Then, on $H$ it is natural to choose
$\left(\frac{\partial}{\partial v}\right)^a := T^a$.  Note that nowhere in
this construction have we assumed that  $T^a$ be everywhere timelike.  Just as
the natural (Killing) evolution vector field for the Kerr spacetime is not
everywhere causal (even outside the event horizon), so in this case $T^a$ is
allowed to be null or spacelike.  Regardless of the nature of $T^a$, it can be
decomposed into components normal and tangent to the three-surfaces
$\Sigma_{t}$ as 
\begin{eqnarray}
  T^a = N \hu^a + V^a \, ,  
\end{eqnarray}
where $N$ is the lapse function and $V^a \in T\Sigma_t$ is the shift vector
field.  Given that $T^a$ generates $B$ and $H$ we can perform similar
decompositions on those surfaces. Respectively we have 
\begin{eqnarray} 
  T^a &=& \bN \bu^a + \bV^a \quad  \mbox{ and} \\ 
  T^a &=& \left(\frac{\partial}{\partial v} \right)^{a} 
	= \mathcal{V}^a + \tilde{V}^a  \, , 
\end{eqnarray} 
where $\bN = - T \cdot \bu$, ${V}^a = p_{A}^{\; \; a} \bar{V}^A$ on
$\mathcal{B}_t$ and $V^a = p_{A}^{\; \; a} \tilde{V}^A$  on $S_t$. 
Finally, the foliation parameter can be used to relate the various volume
forms. Specifically, $\hu_a = - N \nabla_a t$, so $\vM = - N \mathbf{dt}
\wedge \vSigma$.  Similarly, on boundaries we have $\vB = \bN \underleftarrow{\mathbf{dt}}
\wedge \vBt$ and $\vH = \underleftarrow{\mathbf{dt}} \wedge \vS$, where 
$\underleftarrow{\mathbf{dt}}$ is the appropriate pullback of $\mathbf{dt}$.  

\subsubsection*{Legendre transform}

\noindent Using this foliation, we can break up the action into space and time components
in the standard way \cite{wald}. 
%
%
%We have introduced a notion of time on the manifold $M$ and its
%boundaries $B$ and $H$ through the introduction of a foliation and an
%evolution vector field $T^a$.  Using this, we can decompose the various terms
%in the action into their spatial and time components in the standard way
%\cite{wald}.  
To this end, we introduce the momentum $P^{ab}$ conjugate to
$h_{ab}$ as
\begin{equation}
  P^{ab} = \frac{1}{16 \pi G} \left(K_{\hat{u}}^{ab} - K_{\hat{u}} h^{ab}
	\right) \, .
\end{equation}
Then, it is straightforward to show that 
\begin{equation}\label{curvature}
  \frac{1}{16 \pi G} N \mathcal{R} = N \mathcal{H} + V^{a} \mathcal{H}_{a} 
  + P^{ab} \Lie_{T} h_{ab} - 2 \Lie_{T} K_{\hat{u}} - K_{\hat{u}} h^{ab}
	\Lie_{T} h_{ab} + 2 D_{a} \left( h^{ab} T^{c} \nabla_{c} \hat{u}_{b} \right)
	\, ,
\end{equation}
where $\mathcal{H}$ and $\mathcal{H}_a$ are the usual Hamiltonian and
diffeomorphism constraints,
\begin{eqnarray}
  \mathcal{H} &=&  \frac{R}{16 \pi G}  
	- 16 \pi G \left( P^{ab} P_{ab} - P^2/2 \right) \label{HamC} 
  \quad \mbox{and} \quad \\
  \mathcal{H}_{a} &=& 2 D_{b} P_{a}^{\; \, b} \label{diffeoC} 
\end{eqnarray} 
which vanish on shell.
 
We can then substitute equation (\ref{curvature}) into our expression for the
action (\ref{IH}) and make use of the relationships between the various
normals at $\mathcal{B}_{t}$ and $S_{t}$ given in (\ref{bubn}) and (\ref{lnu})
to express the action as:
\begin{eqnarray}\label{Legendre}
  I_H & = & \int - dt \left\{ H(t) +
	\int_{\Sigma_t} \mspace{-6mu} \bP^{ij} \Lie_T h_{ij}  
	+ \int_{S_t} \tilde{P}  \Lie_T \vS 
	- \int_{\mathcal{B}_t} \mspace{-8mu} \bar{P} \Lie_T  \vBt
	 \right\}	\, .
\end{eqnarray}
Here the Hamiltonian functional $H(t)$ is given by
\begin{equation}\label{L2}
  H(t)  =  \int_{\Sigma_t} \mspace{-8mu} \vSigma 
	(N \mathcal{H} + V^i \mathcal{H}_i )
	+ \int_{\mathcal{B}_t} \mspace{-8mu} \left\{ 
  \frac{\vBt  \bN  k_{(\bar{n})}}{8 \pi G}  + \vBt \bV^A \bar{\jmath}_A 
  \right\} \, , 
\end{equation}
where 
\begin{equation}
  \bar{\jmath}_A = \frac{p_{A}^{\;\; a} \hu^b\nabla_a \hn_b}{ 8 \pi G} \, .
\end{equation}
Furthermore, 
\begin{equation}\label{bdry_mom}
  \mathbf{P}^{ij} := \vSigma P_{\hu}^{ij} \; , \quad 
  \bar{P} := \frac{\sinh^{-1} \eta}{8 \pi G} \quad \mbox{and} \quad
  \tilde{P} := \frac{\ln\zeta}{8 \pi G}
\end{equation}
are identified as the conjugate momenta to $h_{ij}$, $\vS$, and $\vBt$ in the
usual way \cite{adm, by1, nopaper, bf1}. 

\subsubsection*{The Hamiltonian functional}

\noindent Although it is conventional practice to refer to $H(t)$ as the
Hamiltonian, it is clear that in many cases this functional is not really a
Hamiltonian. The problem is that standard Hamiltonian mechanics properly
applies only to closed systems. Here, however, energy flows are allowed
through the boundaries $\mathcal{B}_t$ and $S_t$ and so in general $\dot{H}(t)
\neq 0$ which in turn means that it cannot really be a Hamiltonian functional.
Technically, the Legendre transform and Hamiltonian identification can only be
correctly applied in those cases where the system is isolated --- that is when
$H$ is an isolated horizon \cite{ih} and $B$ is similarly disconnected from
its surroundings (a sufficient condition is that $\Lie_T \gamma_{ij} = 0$).
However, in \cite{by1} Hamilton-Jacobi arguments were used to demonstrate that
the $H(t)$ derived in this way can still be interpreted as the quasilocal
energy associated with $\Sigma_t$ and its boundaries. These arguments are also
supported by the extended phase space arguments of \cite{bf1}. 

With this in mind, we can consider what $H(t)$ tells us about the quasilocal
energy of the system.  First, since the bulk terms of (\ref{L2}) are pure
constraints, the non-vanishing contributions to the energy come exclusively
from the boundary terms. Thus, it is conventional to identify the value of
each boundary contribution to $H(t)$ as the energy of that boundary. In this
case, the only such term appears on $\mathcal{B}_t$. This term was first
identified in \cite{by1} and has been much discussed since then.  Note,
however, that the value of the boundary energy is not uniquely specified.  At
$B$ we are free to add any free functional of the boundary metric
$\gamma_{ab}$ to the action and hence obtain a different value for the energy
of $B$.  Indeed, in many cases (such as the limit of $\mathcal{B}_t$ going to
infinity in asymptotically flat spacetime) it is necessary to add such
reference terms so that the energy of $B$ doesn't diverge.

Next, we turn our attention to the horizon boundary.  With our choice of
action, we do not obtain a contribution to the Hamiltonian at $H$.  However,
as with the outer boundary $B$, we are free to add any functional of the fixed
data to the action thereby obtaining a different value for the horizon energy.
As we saw in Section \ref{wbh} this is a significant freedom.  
Specifically, if we impose the boundary conditions weakly, we can add
any diffeomorphism invariant function of the fixed boundary data.  The freedom
(\ref{Hterm}) in the action gives rise to a corresponding freedom of
\begin{eqnarray}\label{Haction}
  \Delta H = E[ \tq_{AB}, \tom_A , \lpb, \theta_{(\ell)},v] \, , 
\end{eqnarray}
in $H(t)$.  Thus, as far as the mathematical formalism is concerned the
contribution to the energy from $S_t$ may be any functional of the fixed data.
We will return to a more thorough discussion of the energy associated with a
horizon boundary in Section \ref{energy}.  

First, however,  we re-examine this problem from the
canonical phase space point of view.  This will
both give us a new perspective on the energy calculation and also allow us to
derive an expression for the angular momentum of trapping horizons.

%%%%%%%%%%%%%%%%%%%%%%%%%%%%%%%%
% THE HAMILTONIAN
%%%%%%%%%%%%%%%%%%%%%%%%%%%%%%%%
\section{Canonical phase space formulation}
\label{cpsf}

In the previous section we have seen that trapping horizons can act as
boundaries in a quasilocal action formulation of general relativity. Here we
will see that they are also suitable boundaries in a canonical phase space
description of gravity. Apart from the value that this demonstration has in
its own right, it also gives us another point of view on some of the issues
that arose in the last section.  In particular, it will provide further
elucidation of the relationship between the Hamiltonian and the energy
functional $H(t)$, and a sharpening of the discussion of the allowed freedom
in that functional itself.  Additionally, while the action calculation
permitted us to investigate time evolution and hence obtain the quasilocal
energy functional, this Hamiltonian investigation will allow us to consider
more general evolutions of the boundary data and so also tell us about other
``conserved" quantities. Specifically, we will be able to study angular
momentum as conjugate to rotations of the boundary data.

The particular formalism that we will use was described by the current authors
in a recent paper \cite{bf1} and is designed to deal with situations, such as
this one, where the energy of the quasilocal region is not expected to be
conservered. Such systems cannot be dealt with using standard Hamiltonian
methods which, by their very nature, are tailored to cases where energy is
conserved. The methods that we will use are based on the generalized
Hamiltonian formalisms of classical mechanics which are specifically
designed to deal with situations like these where the energy isn't constant.
These methods expand the phase space to include the time coordinate as a
configuration variable (see for example \cite{arnold}). Then, it is possible
to construct a conserved Hamiltonian that generates time translations in this
new, extended phase space. In this space, the energy of the system is no
longer the value of this Hamiltonian, but instead is the value of the momentum
that is canonically conjugate to time. In \cite{bf1} these methods from
classical mechanics were extended to quasilocal general relativity.%
\footnote{Other applications of extended phase space ideas to general
relativity include the papers listed in \cite{kk}.}  
Just as in classical mechanics, the time coordinate appears as a configuration
variable and the energy as its conjugate momenta.  However, for quasilocal
general relativity, we also include coordinates parameterizing the
two-boundaries in the configuration space.  This allows us to determine the
angular momenta of the boundaries, which are conjugate to rotations of the
boundary data.

\subsection{Phase Space and Symplectic Structure}

The Hamiltonian approach to general relativity is based on a three-dimensional
manifold $\Sigma$ and its attendant fields which evolve in time with respect
to some time parameter $t$, rather than the usual four-dimensional spacetime.
In the case of interest here, $\Sigma$ is diffeomorphic to $[0,1] \times S^2$
and has boundaries $\mathcal{B}$ and $S$ (which are each diffeomorphic to
$S^2$), as was shown in figure \ref{Fig0}.  Over this region, the basic 
defining fields of three-dimensional,
time-dependent general relativity are the three-metric $h_{ij}$ over $\Sigma$
and its canonically conjugate momentum $\bP^{ij} : = \vSigma P^{ij}$.  
These variables contain all the information needed to resconstruct the full
four-geometry  at any time $t$ and, once 
we specify a lapse function $N$ and shift vector $V^i$, also define 
how that geometry evolves in time (Appendix \ref{app2}). 

There are also degrees of freedom on $\mathcal{B}$ and $S$ that determine how 
these boundaries evolve relative to $\Sigma$. To specify these evolutions it is sufficient
to know the ``orthogonal velocity" vector fields --- that is $v_\vdash = (V \cdot \hat{n})/N$.
These degrees of freedom are controlled by the area forms $\vBt$ and $\vS$ and their
conjugate momenta $\bar{P}$ and $\tilde{P}$ which we defined in (\ref{bdry_mom}). 
The encoding of this information is easily seen for $\mathcal{B}$. There, 
\begin{equation}\label{outer_lapse}
  N = \bN \sqrt{1 + \eta^2} \quad \mathrm{and} 
	\quad V^{i} = (\bN \eta) \hat{n}^{i} + \bV^{i} \, , 
\end{equation}
and so $v_\vdash = \eta/\sqrt{1 + \eta^2}$ (recall that $\bar{P}$ is defined in terms of $\eta$). 
On $S$ the relation is a little more complicated, since there are two variables which combine to
define the evolution; $C$ relates horizon evolution to the null vectors and then $\zeta$ ties those
null vectors to $\Sigma$ and $S$. On $S$,
\begin{equation}\label{inner_lapse}
  N = \alpha \left( \zeta - \frac{C}{2 \zeta} \right) \quad \mathrm{and} 
	\quad V^{i} = \alpha \left( \zeta + \frac{C}{2 \zeta} \right) \hn^i +
	\tilde{V}^{i} \, .
\end{equation}
Then $v_\vdash = (2 \zeta^2 + C)/(2 \zeta^2 - C)$ and the evolution in $\Sigma$ 
is given by $\tilde{P}$ (defined in terms of $\zeta$) once we know $C$.

In addition to these configuration and momentum variables,  in the extended
phase space we include the time foliation label $t$ and spatial coordinates on
both boundaries, $\tilde{\theta}^\alpha$ and $\bar{\theta}^{\alpha}$ (for
$\alpha = 1,2$), as configuration variables in order to allow for energy and
angular momentum flows through those boundaries \cite{bf1}.  Along with these
variables come their conjugate momenta $P_t$ (a scalar field) and $\bar{\bP}_\alpha$ and
$\tilde{\bP}_\alpha$ (one-form valued volume forms), which are closely related
to  energy and angular momentum.  We can specify the evolution equations of
the configuration variables; they are simply:
\begin{equation}
  \frac{d}{dt} t = 1 \; , \frac{d}{dt} \bar{\theta}^{\alpha} = 0 \; 
  \mbox{ and } \;  
  \frac{d}{dt}\tilde{\theta}^{\alpha} = 0 \, .
\end{equation}
The values of the conjugate momenta and their evolution equations are
initially left free, however we will see that the necessity of Hamiltonian
evolution will greatly reduce that freedom --- and so generate candidate
expressions for the energy and angular momentum associated with the boundary
$\mathcal{B}$ and horizon $S$.

The allowed values of $(h_{ij}, \bP^{ij}, \vBt, \bar{P}, \vS, \tilde{P}, t,
P_t, \bar{\theta}^{\alpha}, \bar{\bP}_{\alpha}, \tilde{\theta}^{\alpha},
\tilde{\bP}_{\alpha})$ span the phase space of Hamiltonian general relativity
on the manifold $M$.  The canonical symplectic structure%
\footnote{This is really the pre-symplectic structure since the existence of
the Hamiltonian and diffeomorphism constraints means that we have not 
isolated the true degrees of freedom. The proper symplectic structure would
restrict itself to the true degrees of freedom. That said, we will follow the
standard practice and refer to this pre-symplectic structure as the symplectic
structure.} 
on this phase space then takes the form:
\begin{eqnarray}\label{ss}
  \Omega (\delta_1, \delta_2) &=& 
  \int_{\Sigma} \left\{ (\delta_2 \bP^{ij} )(\delta_1 h_{ij}) 
	- (\delta_1 \bP^{ij})(\delta_2 h_{ij}) \right\} 
  + \left\{ (\delta_2 P_t)(\delta_1 t) 
	- (\delta_1 P_t)(\delta_2 t) \right\} \\
  & & + \int_{\mathcal{B}} \left\{ (\delta_2 \bar{P}) (\delta_1 \vBt) 
  - (\delta_1 \bar{P}) (\delta_2 \vBt) \right\}
  +  \int_S \left\{ (\delta_2 \tilde{P})(\delta_1 \vS) 
	- (\delta_1 \tilde{P})(\delta_2 \vS) \right\} \nn \\
  & & + \int_{\mathcal{B}} \left\{ (\delta_2 \bar{\bP}_{\alpha}) 
	(\delta_1 \bar{\theta}^{\alpha}) - (\delta_1 \bar{\bP}_{\alpha})
	(\delta_2 \bar{\theta}^{\alpha}) \right\}
  + \int_{S} \left\{ (\delta_2 \tilde{\bP}_{\alpha})
  (\delta_1 \tilde{\theta}^{\alpha})
  - (\delta_1 \tilde{\bP}_{\alpha})(\delta_2 \tilde{\theta}^{\alpha}) \right\} 
  \, .\nn
\end{eqnarray}

Enforcing boundary conditions will impose restrictions on the allowed
variations on $S$ and $\mathcal{B}$. There, we use the same weak boundary
conditions that we considered in the action calculation in Section \ref{wbh}
as well as imposing the appropriate diffeomorphism constraints. We now
consider how the conditions translate into the canonical framework.  

First, on the outer boundary $\mathcal{B}$, the three-metric
$\gamma_{ij}$ is a combination of  the lapse $\bN$, shift $\bV^{A}$ and two-metric
$\bar{\gamma}_{AB}$.  We fix all of these quantities at the boundary, which in
turn imposes restrictions on the full four-dimensional evolution at the
boundary.  From (\ref{outer_lapse}) we note that in particular this is sufficient to fix
the bulk lapse $N$ and shift $V^i$, up to the freedom encoded in 
$\eta$ (or equivalently $\bar{P}$). 

Next, we consider the inner, horizon, boundary $S$.  The geometry of the
horizon is naturally described in terms of the two-metric $\tilde{q}_{AB}$,
shift vector $\tilde{V}^{A}$, the quantities $\alpha$ and $C$ and other
quantities constructed from the null vector fields $\ell$ and $n$.   These
can, of course, be related to the three-dimensional metric $h_{ij}$, the
momentum $\bP^{ij}$, the lapse and shift, and the quantity $\zeta$.  We will
not give all the relations here, but refer the interested reader to Appendix
\ref{app2}.  In this case the boundary conditions are already tailored to
the two-surface $S$.  They require that we fix the two-metric
$\tilde{q}_{AB}$, the angular momentum one-form $\tilde{\omega}_{A}$ and the
value of $\alpha C$ (which is equivalent to the earlier fixing of $\lpb$), as
well as requiring that $\theta_{(\ell)}$ vanish. 

In analogy to the way that we restricted the variation of these quantities in
the action formulation, here the variation of ``fixed" quantities is 
restricted so that $\delta$ respectively acts as:
\begin{eqnarray}\label{fixedvary}
  \delta_{X_S} = (\delta t) \frac{d}{dt} + \Lie_{\widetilde{\delta X}} 
  \quad \mbox{ and} \quad
  \delta_{X_\mathcal{B}} = (\delta t) \frac{d}{dt} + \Lie_{\overline{\delta X}} \, , 
  \label{X2} 
\end{eqnarray}
on $S$ and $\mathcal{B}$. Here, $\delta t$ is a function of $t$ alone (which
means that the variation preserves the foliation) and $\widetilde{\delta
X}/\overline{\delta X} \in T \mathcal{B} / S$.  Finally, note the difference
between this decomposition of the variation and that given in
(\ref{fixedvary}). In the earlier version, we decomposed with respect to
$\mathcal{V}$ rather than $T$. Thus,  the two are related by: $x_o = \delta t$
and $\widetilde{\delta X}^A = \tilde{x}^A - (\delta t) \tilde{V}^A$.

\subsection{Hamiltonian evolution}

In this section, we will use Hamiltonian methods to obtain expressions for the
energy and angular momentum of a horizon.  In order to do this, we will be
interested in evolutions of the form
\begin{equation}
  \delta_{\Lambda} = \lambda_o \frac{d}{dt} + \Lie_{\lambda} \, , 
\end{equation} 
where $\lambda_o = \lambda_o (t)$ is a scalar function and $\lambda^i \in
T\Sigma$ is a vector field defined over $\Sigma$.  In addition, we require
that $\lambda^i$ be tangent on the boundaries. That is 
\begin{equation}
  \lambda^{i} = p_{A}^{\;\; i} \tilde{\lambda}^{A} \; \mbox{on} \;  S 
  \quad \mbox{and} \quad
  \lambda^{i} = p_{A}^{\;\; i} \bar{\lambda}^{A} \; \mbox{on} \; \mathcal{B} 
  \, , 
\end{equation}
for some $\tilde{\lambda}^A \in TS$ and $\bar{\lambda}^A \in T\mathcal{B}$.
For an evolution along $\Lambda$, we will show that there exists a Hamiltonian
functional which generates the evolution. Specifically, this means that the
symplectic structure (\ref{ss}) with $\delta_{1} = \delta_{\Lambda}$ can be
manipulated into the form 
\begin{equation}\label{Hamform} 
  \Omega(\delta_{\Lambda}, \delta) = \delta K_\Lambda + \mbox{constraints} 
	\, . \end{equation} 
This class of evolutions includes both time evolutions and rotations of the
system, so we can expect the Hamiltonian functional $K_{\Lambda}$ to provide
information about both the energy and the angular momentum associated with the
horizon $S$, subject to an appropriate choice of $\Lambda$.

In order to rewrite the symplectic structure in the form (\ref{Hamform}) we
must evaluate the bulk term by making use of the evolution equations for the
three-metric and extrinsic curvature derived from the Einstein equations.
This is a long calculation and we will simply state the result; we refer the
interested reader to Appendix \ref{app2} for more details.  Before giving the
result, it is useful to introduce a shorthand 
\begin{equation}\label{bulkham}
  H_\Lambda = \int_\Sigma  \left\{ \lambda_o (N \bcH + V^i \bcH_i) 
  + \lambda^i \bcH_i  \right\} \, ,  
\end{equation}
for the bulk part of the Hamiltonian, where $\bcH = \vSigma \mathcal{H}$ and $\bcH_i = \vSigma \mathcal{H}_i$.  As usual this is comprised of
constraints and will therefore vanish on shell.  Then, we can express the bulk
contribution to the symplectic structure as
\begin{eqnarray} 
  \int_{\Sigma} && \left\{ (\delta \bP^{ij} )(\delta_{\Lambda} h_{ij}) 
  - (\delta_{\Lambda} \bP^{ij})(\delta h_{ij}) \right\} \label{bulkssapp} \\ 
  & = & \delta\left( H_{\Lambda}  
  - \int_{\mathcal{B}} \left\{ \frac{\lambda_{o}}{8\pi G} \vBt \bar{N}
    k_{\bar{n}} + \vBt (\lambda_{o} \bar{V}^{A} + \lambda^{A}) \bar{\jmath}_{A}
    \right\} + \frac{1}{8 \pi G} \int_{S}\vS \tilde{\lambda}^{A} \tom_{A}
    \right) \nn \\ 
  && - \int_\Sigma \left\{ \delta(\lambda_o N)
    \bcH + \delta ( \lambda^i + \lambda_o V^i ) \bcH_i \right\} \nn \\ 
  && + \int_{\mathcal{B}} \left\{ (\delta_{\Lambda} \bar{P}) (\delta \vBt) 
    - (\delta \bar{P} ) (\delta_{\Lambda} \vBt) \right\}
    + \int_{S} \left\{ (\delta_{\Lambda} \vS) (\delta \tilde{P})  
    - (\delta \vS)(\delta_{\Lambda} \tilde{P} ) \right\} \nn \\ 
  && + (\delta  \lambda_{o} ) \int_{\mathcal{B}}
    \left\{ \frac{ \bar{N}}{8 \pi G} \vBt k_{\bar{n}} 
    +\vBt \bar{\jmath}_{A} \bar{V}^{A} \right\} 
    + (\delta t) \int_{\mathcal{B}} \Lie_{\Lambda} \left\{ \frac{ \bar{N}}
    {8 \pi G} \vBt k_{\bar{n}} +\vBt \bar{\jmath}_{A} \bar{V}^{A} \right\} 
    \nn \\
  && + \int_{\mathcal{B}} \left\{ (\delta \bar{\lambda}^{\alpha}) \vBt 
    \bar{\jmath}_{\alpha} + (\delta \bar{\theta}^{\alpha}) \delta_{\Lambda}
    (\vBt \bar{\jmath}_{\alpha}) \right\}
    -\frac{1}{8 \pi G} \int_{S} \left\{ (\delta \tilde{\lambda}^{\alpha}) \vS 
    \tom_{\alpha} 
    + (\delta \tilde{\theta}^{\alpha}) \delta_{\Lambda}
    (\vS \tilde{\omega}_{\alpha}) \right\}  \, . \nn
\end{eqnarray}
In order to obtain the above, we have made repeated use of the boundary
conditions.  Specifically, we have used $\theta_{(\ell)}=0$ and the fact that
all other ``fixed'' quantities vary as described in (\ref{fixedvary}).
Additionally, we have made use of the diffeomorphism constraints on both the
horizon and outer boundary.  Then, (\ref{bulkssapp}) is close to the form that
we require.  The first term is an exact variation, and will be a contribution
to the Hamiltonian $K_{\Lambda}$, while the second line will vanish provided
the bulk constraints are satisfied.  The third line consists of terms
involving the boundary area forms and their conjugate momenta.  The final two
lines involve additional terms on the boundaries.  These terms will eventually
lead to our expressions for the energy and angular momentum of the horizon and
outer boundary.  

To shorten and clarify our expressions we introduce further notation.  First,
we define
\begin{eqnarray}\label{bdryconstraint}
  K_{t} & := & P_t - \int_{\mathcal{B}} \vBt
  \left\{ \frac{1}{8 \pi G}  \bN k_{\bn} 
    +\bV^A \bar{\jmath}_{A} \right\} \, , \\
  \bar{\bL}_{A} & := &  \bar{\bP}_{A} 
    - \vBt \bar{\jmath}_{A} \quad \mbox{and} \quad
  \tilde{\bL}_{A}  := \tilde{\bP}_{A} + \frac{\vS}{8 \pi G} \tom_{A} \nn \, .
\end{eqnarray}
We then substitute (\ref{bulkssapp}) into the symplectic structure (\ref{ss}).
Then,
\begin{eqnarray}\label{omlam}
  \Omega (\delta_\Lambda, \delta) & = & 
  \delta \left( H_\Lambda + \lambda_o K_{t} 
  + \int_{\mathcal{B}} \bar{\lambda}^A \bar{\bL}_A 
  + \int_S \tilde{\lambda}^A \tilde{\bL}_A \right) \\
  & &- \int_\Sigma d^3 x \left\{ \delta (\lambda_o N ) \bcH
  + \delta (\lambda_o V^i + \lambda^i) \bcH_i \right\} \nn \\
  & & - (\delta \lambda_o) K_t 
  - \int_{\mathcal{B}}(\delta \bar{\lambda}^{\alpha}) \bar{\bL}_{\alpha} 
  - \int_S (\delta \tilde{\lambda}^{\alpha}) \tilde{\bL}_{\alpha} \nn \\
  & & - (\delta t) (\delta_\Lambda K_t)    
  - \int_{\mathcal{B}} (\delta \bar{\theta}^{\alpha}) 
  (\delta_\Lambda \bar{\bL}_{\alpha})
  - \int_S (\delta \tilde{\theta}^{\alpha}) (\delta_\Lambda \tilde{\bL}_\alpha) 
  \nn \, . 
\end{eqnarray}
The first term line in (\ref{omlam}) is an exact variation.  Therefore, if we
can argue that the remaining terms vanish, this term will be the Hamiltonina
$K_{\Lambda}$ which generates evolutions along $\Lambda$.  This will be the
case provided the five quantities $\bcH, \bcH_{i}, K_{t}, \bar{\bL}_{\alpha}$
and $\tilde{\bL}_{\alpha}$ are all zero.  In addition, the evolution along
$\Lambda$ of $ K_{t}, \bar{\bL}_{\alpha}$ and $\tilde{\bL}_{\alpha}$ must also
vanish.

The vanishing of $\bcH$ and $\bcH_{i}$ is nothing more than the usual
Hamiltonian and diffeomorphism constraints.  The other conditions restrict the
values of the momenta $P_t$ and $\bP_\alpha$.  Until this point, these
quantities have been entirely free, but now we must require
\begin{eqnarray}
  P_t &=& \int_{\mathcal{B}} \vBt \left\{ \frac{\bN k_{\bn}}{8 \pi G} +
    \bV^A \bar{\jmath}_{A}\right\}  \, , \label{Pt}\\ 
  \bar{\bP}_{\alpha} & = & \vBt \bar{\jmath}_{\alpha} 
  \quad \mbox{and} \quad
  \tilde{\bP}_{\alpha} = - \frac{\vS}{8 \pi G} \tom_\alpha \, . \label{Pa} 
\end{eqnarray}
Next, the fact that the $\Lambda$ variation of $K_{t}$ and $\bL_{\alpha}$
vanishes fixes the evolution of the momenta.  Specifically, it guarantees that
the functional forms given in (\ref{Pt}) and (\ref{Pa}) are preserved in time.
Thus, our evolution will be generated by a Hamiltonian provided that the
standard bulk constraints of general relativity are satisfied and in addition
the conjugate momenta introduced on the boundaries take the form given above
in (\ref{Pt}) and (\ref{Pa}).

Thus, if these equations of motion are imposed in the extended phase space,
$\delta_\Lambda$ evolutions are generated by the Hamiltonian
\begin{eqnarray}
  K_\Lambda = H_\Lambda + \lambda_o K_t 
  + \int_{\mathcal{B}} \vBt  \bar{\lambda}^A \bar{\bL}_{A} 
  + \int_S \vS \tilde{\lambda}^A \tilde{\bL}_{A} \, .
\end{eqnarray}
As always for a Hamiltonian, this is constant on-shell. In fact it vanishes
since it is composed entirely of constraints.  However, in the extended phase
space, this does not mean that the energy and angular momenta are identically
zero. Instead, the energy is given by the value of the momentum canonically
conjugate to the time variable. Thus, the formalism tells us that the energy
associated with the evolution $\frac{d}{dt}$ is 
\begin{eqnarray}
  E[d/dt] =  \int_{\mathcal{B}} \vBt 
  \left\{ \frac{ \bN k_{\bn}}{8 \pi G} + \bV^A \bar{\jmath}_A \right\} \, .
\end{eqnarray}
The value of energy we have obtained from the phase space calculation agrees
with the one arising from the Legendre transform in the previous section.
Again, we see that there is no horizon contribution.  Similarly, the angular
momenta of the boundaries are determined by the momenta $\bP_{\alpha}$.
Therefore, the angular momentum $J[\phi]$ associated with a vector field
$\phi^A$ at the two boundaries is
\begin{eqnarray}\label{HamAngMom}
  J_{\mathcal{B}}[\phi] = \int_{\mathcal{B}}  \vBt \phi^{A} 
  \bar{\jmath}_{A} \quad \mathrm{and} \quad
  J_{S}[\phi] = - \frac{1}{8 \pi G} \int_S \vS \phi^A \tom_A \, .
 \end{eqnarray}
We thus obtain angular momentum values for both boundaries independently.  The
value at the outer boundary agrees with what is generally obtained.  The
expression for the angular momentum of the horizon provides an extension of
the previous definition (\ref{ang_mom}) on an isolated horizon.  This phase
space calculation tells us that we can associate an angular momentum with any
vector field $\phi$ at the horizon, not just axisymmetric Killing vectors. As
expected, the value of the angular momentum is determined by the field
$\tom_{A}$.  

\subsubsection*{Freedom in the Hamiltonian}

\noindent Just as there was a freedom in the definition of the action $I_H$
and the quasilocal energy $H(t)$, there is a corresponding freedom in the
definition of our Hamiltonian $K_\Lambda$.  A close examination of
(\ref{omlam}) reveals that $K_\Lambda$ is not the only Hamiltonian that
generates the evolution $\delta_\Lambda$ through the physical phase space.
Consider for example
\begin{eqnarray}
K'_\Lambda = K_\Lambda + \lambda_o F(t) \, , 
\end{eqnarray}
where $F(t)$ is a free function. Then, it is fairly easy to see that
this functional also generates the usual physical evolutions subject to a
modification of the constraint equation (\ref{Pt}) so that it becomes 
\begin{eqnarray}
  P_t &=& \int_{\mathcal{B}} \vBt \left\{ \frac{\bN k_{\bn}}{8 \pi G} +
    \bV^A \bar{\jmath}_{A}\right\} + F(t) \, .
\end{eqnarray}
In the extended phase space, this corresponds to a displacement to an
alternate constraint surface that still corresponds to the usual physical
evolution of the metric and extrinsic curvature. In fact one can argue that
this is the most general term that can be added to the Hamiltonian \cite{bf1}.
With this alternate Hamiltonian, the quasilocal energy becomes 
\begin{eqnarray}
  E[d/dt] = F(t) + 
    \int_{\mathcal{B}} \vBt \left\{ \frac{\bN k_{\bn}}{8 \pi G} +
    \bV^A \bar{\jmath}_{A}\right\} \, .
\end{eqnarray}
The quasilocal angular momentum $J_\phi$ remains unchanged. 

Mathematically, $F(t)$ can take any form but from a physical point of view the
only functionals that make sense are those constructed from from the time
dependent boundary data. That is
\begin{eqnarray}
  F(t) = F[\bar{\gamma}_{AB}, \bN, \bV; \tq_{AB}, \tom_A, \alpha C, 
  \theta_{(\ell)}] \, , 
\end{eqnarray}
and so the quasilocal energy derived by this method is compatible with that
(\ref{Haction}) found by action arguments.

%%%%%%%%%%%%%%%%%%%%%%%%%%%%%%%%
%  ENERGY + ANG MOM
%%%%%%%%%%%%%%%%%%%%%%%%%%%%%%%%
\section{Energy and Angular Momentum of the Horizon}
\label{energy}

In Sections \ref{actSect} and \ref{cpsf} we have obtained expressions for the
energy and angular momentum of a spacetime containing a horizon as a
boundary.  In this section we will consider these expressions in
more detail. In particular we will show that the expression for angular
momentum is equivalent to the corresponding expressions on isolated and
dynamical horizons. Turning to the energy we will see that although a wide
range of energy expressions are allowed by our mathematical formalism, this
freedom may be greatly reduced through the imposition of a few natural
conditions. Once this is done, the remaining freedom is essentially that found
in the dynamical horizon formalism.

\subsection{Angular Momentum}
\label{angmomsect}

The angular momentum of the horizon, derived using the Hamiltonian framework,
is given by
\begin{equation}\label{JpH}
  J_{S}[\phi] = - \frac{1}{8 \pi G} \int_S \vS \phi^A \tom_A \, .
\end{equation}
This expression is unique, as the freedom we have to add terms to the
Hamiltonian $K_{\Lambda}$ will not affect the value of the angular momentum
obtained.  Furthermore, this value of angular momentum is applicable for both
isolated and dynamical horizons.  It is immediate from (\ref{ang_mom}) that
our expression is identical to the standard one on an isolated horizon.

The dynamical horizon definition for angular momentum (\ref{ang_mom_dh}) is
\begin{equation}\label{JAK}
  J_{DH}[\phi] =  \frac{1}{8 \pi G} \int_S \vS \phi^a \hr^b \nabla_a \htau_b 
  \, .
\end{equation}
This is clearly very similar to the expression that we
derived by Hamiltonian methods, but it isn't identical --- (\ref{JpH}) 
made use of the null normals $\ell^a$ and $n^a$ whereas (\ref{JAK}) uses the
spacelike normals $\hr^a$ and $\htau^a$. We will now show that for reasonable
choices of $\phi^A$, this difference is irrelevant and either choice of
normals will give the same value for the angular momentum (up to an overall
sign) associated with $\phi^A$ on the horizon. 

To see this, let us consider the conditions that $\phi^A$ should meet so that
it may be viewed as a generator of rotations. We start with the example of a
two-dimensional slice of the Kerr horizon.  There, the obvious choice for
$\phi^A$ is the Killing vector field, $({\partial}/{\partial \phi})$ in the
usual coordinate systems. Note that in addition to satisfying
$\Lie_{({\partial}/{\partial \phi})} \tq_{AB} = 0$, this vector field can also
be integrated to foliate the two-slice into a family of closed curves, all of
which  have a common period. 

With this example in mind we will assert that $\phi^A$ is a suitable
``rotation" vector field if :
\begin{enumerate}

\item It integrates to a congruence of closed curves that foliate $S$
(rotations should be periodic!).  

\item $\Lie_{\phi} \vS = 0 \Leftrightarrow d_A \phi^A = 0$. Given the first
condition, this one may always be met through a rescaling of $\phi^A$.
Specifically, $\phi^A$ is restricted to take the form
\begin{equation}\label{dp} 
  \phi^A = \beta \tilde{\epsilon}^{AB} d_B \theta \, , 
\end{equation}
where $\theta$ is a labelling of the congruence and $\beta =
\beta(\theta)$ is a free function.  This is a necessary condition for
$\phi^A$ to be a Killing vector field. It may be thought of a requiring a kind
of uniformity of $\phi^A$ along each curve. 

\item Finally, $\beta$ should  be chosen so that the curves all integrate to
have a common period (rotations should have a common period for all orbits!). 
\end{enumerate}
A $\phi^A$ of this type satisfies many, but obviously not all, of the properties of a
Killing vector that generates rotational symmetries.  It is not hard to show
that these conditions are more than sufficient to force the two angular
momentum expressions to be equivalent. Begin by noting that on a (spacelike)
dynamical horizon,
\begin{equation}
  \ell^a = \sqrt{\frac{C}{2}} (\htau^a + \hr^a) \quad \mbox{and} 
  \quad n^a = \frac{1}{\sqrt{2C}} (\htau^a - \hr^a) \, .
\end{equation}
Whence, it easily follows that
\begin{eqnarray}
  \int_{S} \vS \phi^A \tom_A & = & \int_{S} \vS 
  \left\{\phi^a \hr_b \nabla_a \htau^b + \frac{1}{2} \phi^A d_A (\ln C)  
  \right\} \\
  & = & \int_{S} \vS \left\{\phi^a \hr_b \nabla_a \htau^b 
  + \frac{1}{2} d_A (\phi^{A} \ln C) - \frac{1}{2} \ln C \, d_A \phi^A  
  \right\} \, . \nn
\end{eqnarray}
The second term on the second line vanishes because $S$ is closed and the
third term vanishes since $\phi^A$ is required to be divergence-free. 

Thus, with $\phi^A$ of the class that we have discussed, the angular momentum
expression that we derived by Hamiltonian arguments (\ref{JpH}) agrees exactly
with the expression that appears in the Ashtekar--Krishnan flux law
(\ref{JAK}). The two calculations then support each other. Our derivation gives
a Hamiltonian justification for viewing this expression as representing angular
momentum, while the flux law then shows how to calculate changes in that
angular momentum due to matter and energy crossing the horizon.

There are a couple of points to keep in mind with respect to these expressions.
First, the derivation of the dynamical horizon flux law does not actually make use of the 
fact that $H$ is a dynamical horizon, 
except for assuming it is spacelike (so that $\hr^a$ and $\htau^a$ are
defined). Thus, it is true for all spacelike surfaces, not just horizons.
Second, although \cite{ak} is the first paper to derive a flux law of this
type for spacelike surfaces, very similar angular momentum definitions and
flux laws have appeared for timelike and null surfaces.  In fact, while
expressions for quasilocal energy vary widely, angular momentum is almost
always defined with respect to the connection on the normal bundle to a
two-surface. Examples of these expressions include: the Komar formula for
stationary spacetimes (see, for example, \cite{eric}), the definition of
angular momentum for an isolated horizon \cite{ih}, and the Brown-York
definition for quasilocal angular momentum \cite{by1}.  The only difference
between the various definitions is which normal vectors to $S$ are used in the
angular momentum expression. 

\subsection{Energy}

Both the action and Hamiltonian calculations allow for the horizon energy to
be any free functional of the fixed data.  That is,
\begin{equation}
  \mbox{Energy} = E [ \tilde{q}_{AB}, \tilde{\omega}_{A}, \theta_{(\ell)}, 
  \lpb , v] \, , 
\end{equation}
where the exact form of $E$ is unspecified. However, it is also clear that
many of these functionals will not be physically acceptable. To narrow down
the choices, we impose the following conditions on the energy functional:  
\begin{enumerate}
\item  \textit{Dimensional correctness:} When $c = G = 1$, $E$ should have
units of length.

\item \textit{Constancy in isolation:} E should be constant over isolated
sections of horizon.

\item \textit{Scaling with time:} E should rescale linearly with
reparameterizations of time. 
\end{enumerate}
All three of these are very natural. In particular any functional that doesn't
satisfy the first two conditions could hardly be considered an energy. The
third is a little less precise, but is also very fundamental.  Energy is
conjugate to time and so its value should change with rescalings of time.
However, what has been left imprecise in this statement is exactly what we
mean by time. We will now consider this point. 

On the outer boundary $B$, the evolution vector field $T^a$ is naturally
identified with the flow of time. There, $T^a$ appears in the energy
functional
\begin{equation}
  E_{\mathcal{B}} = \int_{\mathcal{B}} \vS \left( 
  \frac{\bar{N} k_{\bu}}{8 \pi} + \bV^A \bar{\jmath}_A \right)
  =  \int_{\mathcal{B}} \vS \, T^a \left( \frac{- k_{\bu} \bu_a }{8 \pi}  
  + \bar{\jmath}_a \right)  \, .  
\end{equation}
Clearly this is dimensionally correct and also rescales with $T^a$. Further,
since $T^a$ is fixed data, any extra reference terms that are added may also
be given the correct scaling behaviour.  Given this experience, on $H$ the
obvious first candidate to consider as a time vector is $(\partial/\partial
v)$ or possibly $\mathcal{V}$. However, with the horizon boundary conditions,
this cannot be the correct choice as these vectors neither appear in our base
expression (which is $0$ on $H$) nor in the fixed data. The only way that we
could get a rescaling with respect to these would be through the foliation
parameter $v$.  However, an expression of this type would clearly fall afoul
of condition 2 and not be constant on isolated horizons. 

Thus, the energy associated with the surface $S$ cannot rescale with either
$(\partial/\partial v)$ or $\cV$.  However, among the fixed data we do have
$\lpb = \alpha C dv$. We are then naturally lead to consider an energy
functional that has a linear dependence on $\lpb$.  Furthermore, the null
vector field $\ell^a$ is causal which gives it an additional attraction over
other, possibly acausal, alternatives. In considering possible forms of the
energy functional, we note that over isolated regions where $\ell_a$ is normal
to $H$, $\lpb = 0$.  Therefore, we propose an energy functional of the
following form:
\begin{equation}
  d E = \frac{1}{c_{o} a} \int_{S} \lpb \wedge \vS \, , 
\label{dE_eq}
\end{equation}
where $a$ is the area of $S$ and $c_{o}$ is a constant.  That is, we argue
that the \textit{energy flux} should be proportional to the ``average" value
of $\lpb$.%
\footnote{Note that our freedoms would equally well allow us to multiply the
right-hand side of this equation by any dimensionless functional $f[\tq_{AB}, \tom_A, v]$ of
the remaining fixed data. However, since $\lpb$ may already be freely chosen,
this does not add to the generality of the expression. }
Then, given cross-sections $S_1$ and $S_2$  which bound a region $\Delta H$
their energies are related by:
\begin{equation}\label{energy_diff_eq1} 
  E_{2} - E_{1} =  \int dv \left\{  \frac{1}{c_{o} a} \int_{S} \lpb \wedge \vS
  \right\}
  = \frac{1}{c_{o}} \int _{v_1}^{v_2}  \overline{\alpha C} dv\, , 
  \end{equation}
where $\overline {\alpha C} = \frac{1}{a} \int_{S} \vS \alpha C$ is the
average value of $\alpha C$ over any surface. Thus, for any choice of $\ell$,
an energy defined in this way will meet our minimum requirements: it will be
dimensionally correct, constant when $H$ is isolated, and (its flux) will
rescale with respect to a causal ``time'' vector field.  However, the energy
in (\ref{energy_diff_eq1}) does not fit naturally into the allowed freedoms of
the action formulation --- we would have to define $E$ as an integral of the
fixed data from some reference surface $S_o$ with energy $E_o$ to the surface
$S$ --- and is even less natural in the Hamiltonian framework.  Therefore, we
will further restrict the freedom in $\lpb$, and hence the form of the
energy.  

For a specific horizon geometry, we would like to fix a unique
normalization for $\lpb$ and consequently a preferred value for the energy.
This is familiar from, for example, the Kerr isolated horizon.  There we
obtain a family of permissible evolution vector fields $\ell$, but restrict to
a preferred one by requiring it to have a specific value of surface gravity.
To make a similar fix here, we need to tie $\lpb$ to the geometry of the
horizon. If the horizon does not contain any ``partially-isolated" two-surfaces
\footnote{That is, two-surfaces which contain both regions where $\lpb = 0$ and 
others where $\lpb \neq 0$.} then a nice way to do this is to require that $\lpb$
satisfy
\begin{equation}\label{ellfix}
   \lpb = c_{o} d f[ \tq_{AB}, \tom_A]  \quad \Leftrightarrow \quad
   \alpha C =  c_{o} \left( \frac{df}{dv} \right)\, ,
\end{equation}
where $f = f[ \tq_{AB}, \tom_A]$ is some functional of the fixed geometric
data on each two-surface $S$. Recently it has been shown 
that a wide class of $\tl = 0$ three-surfaces actually meet this assumption \cite{ams}. 
Roughly speaking, if there are trapped surfaces arbitrarily close to (though spacelike-separated from) 
each two-surface then on those ($\tl =0$) surfaces $\lpb$ vanishes either everywhere or nowhere. 
Thus, for most $H$ which can be identified as black hole boundaries,
one should be able to fix $\lpb$ in this way. 

%
%
%\footnote{This is only possible for horizons which are 

%
%This assumption imposes a restriction on the geometry of the
%horizon by precluding horizons which are ``partially isolated'' in the sense
%that 

%

%that part of a given two surface is isolated (whence $\lpb = 0$) while the rest
%is not. 

% There is no a priori reason to exclude such horizons, although at
%this time there are no known examples.}  
% 
With such a fix, $\lpb$ is ``constant'' over each leaf of the foliation. Then, consulting
(\ref{dE_eq}) it is not hard to see that this 
will uniquely determine the flux of energy through the horizon between
cross-sections --- and so the actual energy expression up to the usual additive
constant.  However, in general relativity the only dimensionful constants are
the speed of light $c$ and Newton's constant $G$.  From these, it is impossible
to construct a quantity with units of energy whence it is natural to set the
additive constant to zero.  Then, we obtain 
\begin{equation} 
 E =  f [ \tq_{AB}, \tom_A ] \, .  
\end{equation}
Thus, with this choice the energy is fixed directly in terms of the geometry
and $\lpb$ is fixed up to an overall constant rescaling.

The energy expression derived above has another interesting property --- it
must always be non-decreasing.  This follows immediately from the fact that the
pull-back of the forward-in-time pointing $\ell$ will always be positive on a
dynamical horizon $H$ (or zero in isolation).  Then, with Kerr black holes and
the Penrose process in mind, this suggests that we associate $E$ with the
irreducible mass of the horizon (see for example \cite{wald}) rather than the
``total" mass which can either increase or decrease in time.  That is we define
\begin{equation}
  f = \frac{R}{2} \, ,
\end{equation}
where $R$ is the areal radius defined by $a =: 4\pi R^{2}$. 

If one wished to instead consider the ``total'' energy it would be necessary to
consider a more complicated evolution vector field. Specifically in the case of
axisymmetry, it would be natural to consider $\ell + \Omega \phi$, where $\phi$
is the rotational Killing vector field and $\Omega$ is a function of the fixed
data. Then the ``irreducible" energy would remain unchanged since
$\underleftarrow{ \mbox{\boldmath{$\ell + \Omega \phi$}} } \wedge \vS = \lpb
\wedge \vS$. However, $\phi^A$ would enter the fixed data as a symmetry of the
two-metric, and so one could also include it in energy expressions.  In
particular, with this symmetry there would be an unambiguous angular momentum
$J$ for each cross-section of the horizon and this could be used to construct
energy expressions like the usual Smarr formula  $E(a,J)$ for the total mass of
a Kerr black hole.

\subsubsection*{Relation to dynamical horizons}

We will now show that the energy obtained by Hamiltonian methods is related to
the dynamical horizon energy and flux laws.  Furthermore, by appealing to
dynamical horizons, we can fix the unknown constant $c_{o}$ and hence further restrict
the energy.  We begin by reviewing those aspects of the dynamical horizon
calculation which are relevant here.  

Dynamical horizons are required to be spacelike, therefore, we can introduce
the unit timelike and spacelike normals $\hat{\tau}^{a}$ and $\hat{r}^{a}$.
From these, we obtain a null vector 
\begin{equation}\label{hat_l}
  \hat{\ell}^{a} = \hat{\tau}^{a} + \hat{r}^{a} \, .
\end{equation}
The dynamical horizon flux law is obtained by integrating a
combination of the Hamiltonian and diffeomorphism constraints over the
horizon.  The specific combination is chosen to correspond to the
$\hat{\ell}^{a} \hat{\tau}^{b}$ component of the Einstein equations.  Then,
by making use of the fact that at the horizon $\theta_{(\ell)} = 0$, one
obtains an energy flux law
\begin{equation}\label{flux}
  E_{2} - E_{1} = \int_H \vH T_{ab} \xi_{E}^a \htau^b 
  + \frac{1}{16 \pi G} \int_H \vH N_{E} \left\{ \sigma_{(\ell)}^{AB} 
  \sigma_{(\ell) AB} + 2 \zeta^A \zeta_A \right\} \, .  
\end{equation}
In the above, $\vH$ is the horizon volume form, $\zeta_{A} = p_{A}^{a}
\hat{r}^b \nabla_b \hat{\ell}_a$, $N_{E}$ is the lapse on the horizon and
$\xi_{E}^{a}$ is the evolution vector field associated with the energy
functional $E$.  Finally, the energy and evolution vector are tied together
via the lapse by
\begin{equation}\label{lapse_cond}
  N_{E} = 2 || dE || \quad \mbox{and} \quad
  \xi_{E}^{a} = N_{E} \hat{\ell}^{a} \, .  
\end{equation}
The expression (\ref{flux}) provides a nice relationship between the change in
energy of a horizon and the flux of gravitational and matter energy through
the horizon.  The change in energy is shown to be equal to three terms, all of
which are non-negative (provided the weak energy condition is satisfied).  The
first is identified as the flux of matter through the horizon and the
second is the flux of gravitational shear.  There is no clear interpretation
of the third contribution.

Thus, the dynamical horizon energy always increases in the same way as the
``irreducible'' energy that we have derived.  Further, we note that
(\ref{lapse_cond}) can be rewritten without reference to unit vectors (with
the help of (\ref{hat_l})) as
\begin{equation}\label{pbxi} 
  \underleftarrow{\bxi} = 2 d E \, .
\end{equation}
This suggests that we fix the unknown constant arising in (\ref{ellfix}) as
$c_{o} = 1/2$.  Then, the Hamiltonian evolution vector field $\ell^{a}$ is equal
to the evolution vector $\xi^{a}$ arising in the flux law (\ref{flux}).
Therefore, the Hamiltonian evolution vector field and energy also satisfy the
dynamical horizon flux law.

In summary, a simple argument has lead us to obtain an energy functional for
the horizon.   This energy is constant whenever the horizon is isolated.
Additionally, by arguing that the energy should rescale appropriately with
reparametrizations of time, we have identified the null vector $\ell$ as the
evolution vector field of the horizon.  This is a surprising result.  One would
usually expect the evolution vector field to be tangent to the boundary, but in
this case it will not be, unless the horizon is isolated.  Next, by relating
the value of $\ell$ to the other geometrical properties of the horizon, we
obtain a energy associated to $\ell$ which is unique up to an overall constant
rescaling. It is non-decreasing and so naturally associated with the
irreducible (rather than total) mass of the horizon. Finally, by appealing to
the dynamical horizon results (which also describe non-decreasing energies), we
can restrict the little remaining freedom in the energy in such a way that
$\ell$ and $E_{\ell}$ satisfy the dynamical horizon flux laws.

%%%%%%%%%%%%%%%%%%%%%%%%%%%%%%%%
% DISCUSSION
%%%%%%%%%%%%%%%%%%%%%%%%%%%%%%%%

\section{Discussion}

There are many different local definitions for black hole horizons: apparent,
trapping, dynamical, isolated, and slowly evolving.  Although not identical, these
all share some common features.  Specifically, they are all three-surfaces 
which can be foliated by two-surfaces of zero outward expansion.
More precisely, for each two-surface the outward directed null normal $\ell$
to the surface has zero expansion, $\tl = 0$.  In this paper, we presented
both action and canonical phase space formulations for a spacetime with such a 
boundary.  These formulations provide us with quasi-local expressions
for the horizon energy and angular momentum.

A trapping horizon has one natural boundary condition, namely that the outward
expansion $\tl$ vanishes.  This in itself is a mixed boundary condition in the
sense that it depends upon both the intrinsic and extrinsic geometry of the
surface.  We impose additional boundary conditions which fix the two-metric
$\tq_{AB}$ and the one-form $\tom_A$ on every slice of the horizon.  The one-form 
$\tom_{A}$ is the connection on the normal bundle to the horizon.
Physically, these fix the intrinsic geometry of the two-surfaces as well as
their associated angular momenta.  Unlike the $\theta_{(\ell)} = 0$ condition
which holds for all rescalings of the null vector $\ell$, the one-form
$\tom_{A}$ depends upon the particular normalization of $\ell$.  Thus, we must
also fix a preferred normalization for $\ell$, and this is done by fixing the
pull-back of $\ell$ to the horizon. 

We have shown that these are suitable boundary conditions for both the action
and Hamiltonian formulations of general relativity.  While this is an interesting
result in its own right, in this paper we are mostly interested in the expressions that we can
then derive for the
energy and angular momentum associated with
the horizon two-surfaces.  The result for the angular momentum is unambiguous:
there is a definite value of angular momentum associated with each cross
section of the horizon.  Furthermore, this value is identical to the angular
momentum calculated for axisymmetric isolated horizons.  In addition, if we
restrict attention to divergence free vector fields, the value of the angular
momentum agrees with that arising in the dynamical horizon flux law.

Obtaining an energy for the horizon is not so straightforward.  As in all
quasilocal Hamiltonian calculations, there is a freedom to add a functional of
the fixed boundary data to the Hamiltonian and hence change the value of the
energy.  However, we can almost entirely remove this freedom by requiring that
the energy satisfy a few simple physical properties. These then lead  to
several interesting results.  The first is that the appropriate ``evolution
vector'' for the horizon is the null vector $\ell$ and hence, in contrast to a
typical Hamiltonian formulation, the evolution vector need not map cross
sections of the horizon into each other.  Next, by requiring that the null
vector $\ell$, or more accurately its pull-back to the horizon, is compatible
with the horizon geometry we obtain a range of possible expressions for the
energy of the horizon: For a particular choice of null vector field $\ell$ the
horizon energy is fixed up to an overall constant rescaling. All of these
energies are constant on isolated horizons and increasing on dynamical
horizons, and can therefore be interpreted as describing the irreducible mass
of the horizon.  Finally, appealing to the dynamical horizon formulation we can
then fix the value of the rescaling constant relating the energy to the
evolution vector $\ell$.  Once this is done, we find that our energy $E_{\ell}$
and evolution vector field $\ell$ satisfy the dynamical horizon flux law. 

Note however that this is certainly not meant to imply that all problems concerning energy and 
dynamical horizons have been resolved. While it is remarkable that the allowed energy expressions
obtained by these two very different methods agree, it is also important to keep in mind that 
neither method singles out a unique expression. While this is somewhat disappointing it is also not
too surprising given the well-known ambiguities of gravitational energy. 

Other problems also remain for future investigations. Probably the most obvious arises on recalling
the discussion of footnote \ref{two}. There it was pointed out that these $\tl = 0$ horizons
are not uniquely defined --- in most spacetimes a given horizon will be part of an entire family of $\tl = 0$
three-surfaces which may be smoothly deformed into each other. Thus, in using these techniques 
to discuss energy flows, one needs to be careful as quantities
such as cross-sectional area will certainly depend on the member of the family chosen. This 
is not necessarily a problem. In physics, it is common for observations to vary with reference frames
(think special relativity). The important thing is that invariant physical predictions should be compatible.
Thus, for example, while the total energy absorbed by a black hole during a finite incidence of gravitational waves should be invariant, the details of how that happens may depend on the
specific ``horizon". 
 
Lastly, the formalism and boundary conditions presented here may find other
applications. Perhaps the most interesting is the possibility that the
boundary conditions that we have discussed could find a use in quantum gravity
discussions of dynamic black holes --- just as the isolated horizon boundary
conditions and Hamiltonian formalism found an application in the loop quantum
gravity computations of black hole entropy. 

\section*{Acknowledgements}

We would like to thank Abhay Ashtekar, Christopher Beetle, Jolien Creighton
and Badri Krishnan for useful discussions and also acknowledge the careful reading   
and constructive suggestions of our referees. S.F. was supported by the Killam
Trusts at the University of Alberta and NSF Grant No PHY-0200852; I.B.  was
supported by NSERC.

%%%%%%%%%%%%%%%%%%%%%%%%%%%%%%%%
%  APPENDICES
%%%%%%%%%%%%%%%%%%%%%%%%%%%%%%%%

\appendix

\section{Details of the action variation}
\label{app1}

In this section we show that
\begin{eqnarray}\label{actionvar}
  \delta I_H &=& 
	\frac{1}{16 \pi G} \int_M \mspace{-10mu} \vM G_{ab} \delta g^{ab} 
	- \frac{1}{16 \pi G} \int_{\Sigma_2-\Sigma_1} \mspace{-36mu} \vSigma 
	\left( K_{\hu}^{ij} - K_{\hu} h^{ij} \right) \delta h_{ij} \\ 
	& & + \frac{1}{16 \pi G} \int_B \mspace{-6mu} \vB 
	\left( K_{\bar{n}}^{ij} - K_{\bar{n}} \gamma^{ij} \right) \delta \gamma_{ij} 
	+ \frac{1}{8 \pi G} \int_{\mathcal{B}_2-\mathcal{B}_1} \mspace{-32mu}
		 (\delta \vBt) \sinh^{-1} (\eta) \nonumber \\ 
	& & + \frac{1}{16 \pi G} \int_H \mspace{-6mu} \bdv \wedge \vS  
	\{\tilde{s}^{AB} \delta \tq_{AB} \} 
	+ \frac{1}{8 \pi G} \int_{S_2-S_1} \mspace{-32mu} (\delta \vS )
	\left( 1 - \ln \zeta \right) \nn \\ 
	& &  + \frac{1}{8 \pi G} \int_H  \mspace{-6mu} \left\{ \vS \wedge \left(
	(\delta \lpb) \theta_{(n)} - (\delta \npb) \theta_{(\ell)} \right) 
	+ \tilde{V}^A \,\delta \left(\bdv \wedge \vS \tilde{\omega}_{A} \right) 
	- \delta \left( 2 \npb \theta_{(\ell)} \wedge \vS \right) \right\} \nn \, .  
\end{eqnarray}
as asserted at equation (\ref{dIH}). In the interests of brevity, the
(well-known) outer boundary terms on $B$ will be omitted from this derivation.
Details of the calculations for those terms can be found in a standard
reference such as \cite{wald, eric}. 

We begin by writing the bulk term in the action as:
\begin{equation}
  \delta \left( \vM \mathcal{R} \right) = 
	\vM \left\{ G_{ab} \delta g^{ab} + \nabla^{a} \left[ g^{bc} 
	\left( \nabla_{b}	\delta g_{ac}  - \nabla_{a} \delta g_{bc}  \right) 
	\right]\right\}
\end{equation}
On integrating over $M$ and using Stokes theorem to move total derivatives out
to the boundaries this becomes:
\begin{eqnarray}\label{v1} 
  \delta \left( \int_M \mspace{-6mu} \vM \mathcal{R} \right) &=& 
	\int_M \mspace{-6mu} \vM G_{ab} \delta g^{ab} 
  + \int_{\Sigma_2-\Sigma_1} \mspace{-32mu} \vSigma \hu^{a} 
	g^{bc} \left( \nabla_{b}	\delta g_{ac}  - \nabla_{a} \delta g_{bc}  \right) 
	\\
  & & + \int_H \mspace{-6mu} \bdv \wedge \vS \tau^a g^{bc} 
	\left( \nabla_{b}	\delta g_{ac}  - \nabla_{a} \delta g_{bc}  \right) \, .
	\nonumber
\end{eqnarray}
In order to obtain the term on the
horizon, we have used the expression (\ref{horizonvol}) for the volume form.  
With just a little algebra, the term on $\Sigma_{1,2}$ can be
written as:
\begin{eqnarray}
  & &\int_{\Sigma_2-\Sigma_1} \mspace{-32mu} \vSigma \hu^{a} g^{bc} 
	\left( \nabla_{b}	\delta g_{ac}  - \nabla_{a} \delta g_{bc}  \right) \\
  & & \qquad = - \int_{\Sigma_2-\Sigma_1} \mspace{-32mu} \left\{ 
	\vSigma (K_{\hu}^{ij} - K_{\hu} h^{ij}) \delta h_{ij} 
	+ 2 \delta ( \vSigma K_{\hu} ) \right\} 
	- \int_{S_2-S_1} \mspace{-26mu} \vS \hn_a \delta \hu^a \, . \nn
\end{eqnarray}
The last term is not, of course, zero since variations of the ``time-space"
components of $g^{ab}$ can give $\delta \hu^a$ a component in the $\hn_a$
direction. This is in contrast to $\delta \hu_a$ whose ``direction", though
not normalization, is fixed by the requirement that it be normal to the
$\Sigma_{1,2}$ surfaces. 

The signature on $H$ is indeterminate, so in analyzing the terms in (\ref{v1})
on that boundary we must avoid the use of the inverse metric $q^{ij}$ as well
as proper volume forms or unit-normalized normal vector/one-form fields. With
these caveats in mind we note that:
\begin{eqnarray}
  \tau^a g^{bc} \left( \nabla_{b}	\delta g_{ac}  
	- \nabla_{a} \delta g_{bc}  \right) = 
	\left(\tau^{a} \tilde{q}^{bc} + n^a \ell^b \mathcal{V}^c \right)  
	\left( \nabla_{b}	\delta g_{ac} 
	- \nabla_{a} \delta g_{bc}  \right) 
\end{eqnarray}
In turn, the two expressions on the right-hand side of this equations can be
rewritten as:
\begin{eqnarray}\label{tauterm}
  &&\tau^{a} \tilde{q}^{bc} \left( \nabla_{b}	\delta g_{ac} 
	- \nabla_{a} \delta g_{bc}  \right)  \\
	&& \qquad = - 2 \delta k_\tau - \delta \tq_{ab} k_\tau^{ab}
	+ \tq^{ab} \nabla_a \delta \tau_b 
  + \tq^{\, a}_b \nabla_a \delta \tau^b
	- (\delta \tilde{q}_{ab}) \tilde{q}^{ac} (\ell^b n^d + n^b \ell^d)
	\nabla_{c} \tau_{d} \, , \nonumber
\end{eqnarray}
where $k_\tau = \tq^{ab} \nabla_a \tau_b$, and
\begin{eqnarray}\label{Vterm}
  & & n^a \ell^b \mathcal{V}^c \left( \nabla_{b} \delta g_{ac} 
	- \nabla_{a} \delta g_{bc} \right) \\
  & & \qquad = - 2 \delta \kappa_{\mathcal{V}} 
	+ 2 (\delta \mathcal{V}^i) \omega_i 
	- \mathcal{V}^b (\delta n_a \nabla_b \ell^a + n_a \nabla_b \delta \ell^a 
	+ \delta n^a \nabla_b \ell_a + n^a \nabla_b \delta \ell_a )\nn \, .
\end{eqnarray}
There is still significant work needed to transform these expressions into
what is present in (\ref{actionvar}).  We begin by considering the right hand
side of equation (\ref{tauterm}).  First, we note that the $\delta k_{\tau}$
term can be rewritten using:
\begin{equation}\label{mess0}
  \vS k_{\tau} = \vS \tilde{q}^{ab} \nabla_{a} \tau_{b} 
	= - \vS \tilde{q}^{ab} \nabla_{a} \mathcal{V}_{b} 
	+ 2 \vS \alpha \theta_{(\ell)} 
	= - \Lie_{\mathcal{V}} \vS + 2 \vS\alpha \theta_{(\ell)} 
\end{equation}
Then the $\Lie_{\mathcal{V}} \vS$ can be integrated out to the future and past
boundaries of the horizon while the second term (and its variation) will
vanish since we require that $\theta_{(\ell)} = 0$ on the horizon.  The second
term in (\ref{tauterm}) appears as a contribution to $\tilde{s}^{AB} \delta \tq_{AB}$ in the
final expression\footnote{To be precise, note that $k_\tau^{ab} \delta \tq_{ab} = - k_{\tau ab} \delta \tq^{ab} =  - k_{\tau ab} \delta (p^{\; a}_A p^{\; b}_B \tq^{AB} )= -  k_{\tau AB}  \delta \tq^{AB}
= k_{\tau}^{AB} \delta q_{AB}$.   }.  Next, we turn to the third and fourth terms in equation
(\ref{tauterm}), those containing a $\delta \tau$.  By expanding $\tau$ in
terms of $\ell$ and $n$ using (\ref{alphaDef}), it follows that they can be
rewritten as
\begin{eqnarray}\label{mess1}
  &&\tq^{ab} \nabla_a \delta \tau_b + \tq^a_b \nabla_a \delta \tau^b \\
	&& \qquad  = 
	2 (\delta \alpha) \theta_{(\ell)} + 2 \delta (\alpha C) \theta_{(n)} 
	+ \tilde{q}^{ab} \nabla_{a} (\tilde{q}_{bc} \delta \tau^c) 
  - (\tilde{q}^{ab} \Lie_{\mathcal{V}} \tilde{q}_{ab}) 
	(n^a \delta \ell_a + n_a \delta \ell^a ) 
	\, . \nonumber
\end{eqnarray}
The first two terms arising above appear in the final formula
(\ref{actionvar}) and the third is exact and will therefore vanish upon
integration over a two surface $S_{v}$.  The final term will combine with
others to form an exact derivative.  To see this, we add the last term on the
right hand side of (\ref{tauterm}) to the last term of (\ref{Vterm}).  By
making use of the fact that $n^a \nabla_{b} \tau_a = n^a \nabla_{b}
\mathcal{V}_a$ and $\ell^a \nabla_{b} \tau_a = - \ell^a \nabla_{b}
\mathcal{V}_a$ it follows that
\begin{eqnarray}\label{mess2}
  &&-(\delta \tilde{q}_{ab}) \tilde{q}^{ac} (\ell^b n^d + n^b \ell^d)
	\nabla_{c} \tau_{d} 
	- \mathcal{V}^b (\delta n_a \nabla_b \ell^a + n_a \nabla_b \delta \ell^a 
	+ \delta n^a \nabla_b \ell_a + n^a \nabla_b \delta \ell_a ) \nonumber \\ 
	&&\qquad = -\Lie_{\mathcal{V}} (n^a  \delta \ell_a + n_a \delta \ell^a) \, .
\end{eqnarray}
When added to the $\Lie_{\mathcal{V}}$ term from (\ref{mess1}) we obtain an
exact derivative which can be integrated out to the future and past horizon
surfaces $S_{1}$ and $S_{2}$.  

Finally, we turn our attention to the expressions in (\ref{Vterm}).  The first
appears in the $\tilde{s}^{AB} \delta \tq_{AB}$ 
of the final expression 
(through $\tq^{AB} \delta \tq_{AB} = \tq^{ab} \delta \tq_{ab}$)
while the last was
dealt with in (\ref{mess2}).  Therefore, we are left only with the $2 (\delta
\mathcal{V}^i) \omega_i$.  This can be rewritten using $\delta (\mathcal{V}^i
\bdv) \wedge \vS = -p_{A}^{\;\; i}  (\delta \tilde{V}^A) \bdv \wedge \vS$ to
show that
\begin{eqnarray}\label{mess3}
  2 \,\bdv \wedge \vS (\delta \cV^i) \omega_{i} 
	= - 2 \,\delta \left( \bdv\wedge \vS \tilde{V}^{A} \tom_A \right) 
	- 2 (\delta \, \bdv ) \wedge \vS \, \kappa_{\mathcal{V}} 
	+ 2 \, \tilde{V}^{A} \delta ( \bdv \wedge \vS \tom_A ) \, .
\end{eqnarray}

The results from (\ref{mess0} - \ref{mess3}) can be used to rewrite the
horizon terms from (\ref{v1}) as:
\begin{eqnarray}\label{horizonterm}
  \int_H \mspace{-6mu} \bdv \wedge \vS \tau^a g^{bc} 
	\left( \nabla_{b}	(\delta g_{ac}) - \nabla_{a} (\delta g_{bc}) \right)
	&=& - 2 \, \delta \left(\int_H \left\{ \bdv \wedge \vS 
	\left( \kappa_{\mathcal{V}} + \tilde{V}^{A} \tom_{A}\right) 
	+ 2 \npb \theta_{(\ell)} \wedge \vS \right\} \right) \\
	&& + \int_H \bdv \wedge \vS \left[ \left( \tilde{q}^{AB} k_{\tau} 
	-	k_{\tau}^{AB} + \tilde{q}^{AB} \kappa_{\mathcal{V}} \right) \delta
		\tilde{q}_{AB} \right] \nonumber \\
  &&  + \int_H \left\{ 2 (\delta \lpb) \wedge \vS \,  \theta_{(n)} 
	- 2 (\delta \npb) \wedge \vS \,  \theta_{(\ell)} + \tilde{V}^{A}
	  \delta(\bdv \wedge \vS \tom_{A}) \right\} \nonumber \\
	&& + \int_{S_2-S_1} \mspace{-32mu} \left\{ 2\, \delta \vS - \vS \left[
	n_{b} \delta \ell^b + n^{b} \delta \ell_b \right] \right\} \nonumber \, .
\end{eqnarray}
In the above, $(\lpb)_{i} = p_{i}^{\;\; a} \ell_{a} = \alpha C [dv]_{i}$ and
$(\npb)_{i} = p_{i}^{\;\; a} n_{a} = \alpha [dv]_{i}$.  By using the
expression (\ref{tildes}) for $\tilde{s}^{AB}$ we see that contribution on $H$
is of the form arising in (\ref{actionvar}), leaving us only the corner terms
arising at $S_{1}$ and $S_{2}$ to consider.  We have a contribution $n_{b}
\delta \ell^b + n^{b} \delta \ell_b $ arising from $H$ and another
contribution, $\hn_a \delta \hu^a$ from $\Sigma_{1,2}$.  However, there are
relationships between the unit vectors $\hat{u}$, $\hat{n}$ and $\ell$, $n$
given in (\ref{lnu}).  Making use of these, and the fact that $\delta
\hat{u}_a \propto \hat{u}_a$, it follows that
\begin{equation}\label{deltazeta}
  \hat{n}^{a} \delta \hat{u}_{a} + n_{a} \delta \ell^{a} + n^{a} \delta
	\ell_{a} = -2 \delta (\ln \zeta) \, .
\end{equation}
Then, combining (\ref{horizonterm}), (\ref{deltazeta}) and recalling terms
in the action (\ref{IH}) on $\Sigma_{1,2}$ and $H$, it is straightforward to
arrive at the desired result, (\ref{actionvar}).

\section{Manipulating the bulk symplectic structure}
\label{app2}

In this appendix, we will rewrite the bulk symplectic structure in a form
which is directly useful in the body of the paper.  We are interested in the
case where $\delta_{1} = \delta_{\Lambda}$ generates an evolution along the
vector field $\Lambda^{a}$.  In three-dimensional language, the evolution of
the three-metric, $h_{ij}$ and its conjugate momentum $P^{ij}$ which encodes
the extrinsic curvature degrees of freedom are given in by:
\begin{equation}
  \delta_{\Lambda} h_{ij} = \lambda_{o} \frac{d}{dt} h_{ij} 
  + \Lie_{\lambda}  h_{ij} \quad \mbox{and} \quad
  \delta_{\Lambda} P^{ij} = \lambda_{o} \frac{d}{dt} P^{ij} 
  + \Lie_{\lambda}  P^{ij} \, .
\end{equation}
If the bulk fields are required to satisfy the Einstein equations, then we can
easily derive expressions for the time evolution of $h_{ij}$ and $P^{ij}$ as:
\begin{eqnarray} 
  \frac{d}{dt} h_{ij} &=& 64 \pi G N (P_{ij} - \frac{1}{2} P h_{ij} )
  + \Lie_V h_{ij} \, . \label{hdot}  \\ 
  \frac{d}{dt} \bP^{ij} &=& \frac{\vSigma}{16 \pi G} \left( N (R^{ij} 
  - \frac{1}{2} R h^{ij}) - (D^i D^j N - h^{ij} D_k D^k N) \right) 
  + \Lie_V \bP^{ij}  \label{Pdot}\\ 
  & & + 8 \pi G N \left( (P^{kl} \bP_{kl} - \frac{1}{2} P \bP ) h^{ij}
  - 4 ( P^{k (i}P_k^{\; j)} - \frac{1}{2} P \bP^{ij} ) \right) \, , \nn
\end{eqnarray}
where $R$ is the three-dimensional Ricci scalar and $P = h_{ij} P^{ij}$ and
$N$ and $V^{a}$ are the lapse and shift respectively.  Then, a fairly standard
calculation \cite{thesis, BLY} allows us to rewrite the bulk symplectic
structure as:
\begin{eqnarray} 
& &   \int_{\Sigma}  \left\{ (\delta \bP^{ij} )(\delta_{\Lambda} h_{ij}) 
  - (\delta_{\Lambda} \bP^{ij})(\delta h_{ij}) \right\} \label{bulkss} \\ 
  & = & \int_\Sigma
    \left\{ \lambda_o N \delta \bcH + ( \lambda^i + \lambda_o V^i ) \delta
    \bcH_i \right\} \nn \\ 
  & & + \int_{\partial \Sigma} \vS \left\{ - \lambda_o (\hat{n}_{k} V^{k}) 
    \tilde{P}_{CD} \tilde{q}^{CA} \tilde{q}^{DB} 
    + \frac{N \lambda_{o}}{16 \pi G} (k_{\hat{n}} \tilde{q}^{AB} 
    - k_{\hat{n}}^{AB}) 
    + \frac{\lambda_{o} \hat{n}^{i} D_{i} N}{16 \pi G} \tilde{q}^{AB} \right\}
    \delta \tilde{q}_{AB} \nn \\ 
  & & + \int_{\partial \Sigma} 2 \delta (\vS \hat{n}^i P_{ij} p_{A}^{\;\; j} )
    (\lambda_{o} \tilde{V}^{A} + \tilde{\lambda}^{A} )  
    + \int_{\partial \Sigma} \left\{ 2 \lambda_{o} (\hat{n}_{i} V^{i}) 
    \delta (\vS P^{ij} \hat{n}_{i}\hat{n}_{j}) 
    - \frac{N \lambda_{o}}{8 \pi G} \delta ( \vS k_{\hat{n}} ) \right\}\, ,
    \nn 
\end{eqnarray}
where in the interests on conciseness we use tildes to represent
two-dimensional quantities on all boundaries. Thus, $\tilde{P}_{AB} =
p_A^{\;\; i} p_B^{\;\; j} P_{ij}$ is the pull-back of $P_{ij}$, $\tilde{V}^A$
and $\tilde{\lambda}^A$ are the components of $V^i$ and $\lambda^i$ in $T
(\partial \Sigma)$, and $\tilde q_{AB}$ is the two-dimensional metric on
$\partial \Sigma$.  Additionally, since $\hat{n}$ is a unit vector in $\Sigma$
but orthogonal to $\partial \Sigma$, $k_{\hat{n}}^{AB}$ is the extrinsic
curvature of the boundary $\partial \Sigma$ in $\Sigma$.  

Now, we would like to rewrite the boundary contributions in terms of
quantities which are tailored to the boundaries $B$ and $H$.  In the
four-dimensional context, we are more interested in the normals to $B$ and $H$
in $M$ rather than $\hat{n}$, the normal to $\partial \Sigma$ in $\Sigma$.
The former are independent of our spacetime slicing while the latter depends
upon the foliation of the spacetime.  To facilitate this procedure, we
re-express the familiar boundary quantities from Section \ref{actSect} in
terms of the three-dimensional quantites appearing above.  We begin by writing
out the various horizon quantities in terms of the three-dimensional metric
and extrinsic curvature as:

\begin{eqnarray}  
  \tilde{\omega}_{A} &=& (16 \pi G) p_A^{\;\;  i} P_{ij} \hat{n}^{j} 
	+ d_{A} (\ln \zeta) \\
  \kappa_{\mathcal{V}} &=& \frac{d}{dt} ( \ln \zeta ) 
	- \Lie_{\tilde{V}} \ln \zeta 
	+ 16 \pi G (V^{k} n_{k}) \left[P_{ij} n^{i} n^{j} - P^{2}/2 \right]
	+ \hat{n}^i D_{i} N \label{kappa} \\
	\theta_{(\ell)} &=& \zeta \left[ - (16 \pi G) P_{ij} \hat{n}^{i} 
	\hat{n}^{j} + D_{i} \hat{n}^{i} \right] \label{expl} \nn \\
	\theta_{(n)} &=& \frac{-1}{2 \zeta} \left[ (16 \pi G) P_{ij} 
	\hat{n}^{i} \hat{n}^{j} + D_{i} \hat{n}^{i} \right] 
	\label{expn} \\
	k_{\tau}^{AB} &=& (16 \pi G \hat{n}_{i} V^{i}) 
	\tilde{q}^{AC} \tilde{q}^{BD} \tilde{P}_{CD} + N k_{\hat{n}}^{AB} 
  \label{ktau} \\
	\tilde{s}^{AB} &=& - (16 \pi G V^{i} \hat{n}_i) \tilde{q}^{AC} \tilde{q}^{BD}
	\tilde{P}_{CD} + N \left(k_{\hat{n}} \tilde{q}^{AB} - k_{\hat{n}}^{AB} 
  \right) 
	+ (\hat{n}^{i} D_{i} N) \tilde{q}^{AB} \nonumber\\
  &&+ \tilde{q}^{AB}\left[ \frac{d}{dt} ( \ln \zeta )
  - \Lie_{\tilde{V}}(\ln \zeta) \right] \label{sab}
\end{eqnarray}
Similarly, we can write the outer boundary terms as:
\begin{eqnarray}
  k_{\bar{n}}^{AB} &=& \sqrt{1 + \eta^2} k_{\hat{n}}^{AB} 
  + (16 \pi G \eta) \left[ \bar{\gamma}^{AC} \bar{\gamma}^{BD} \tilde{P}_{CD} 
  - P \bar{\gamma}^{AB} /2 \right] \label{knbar} \\
  \bar{n}^{c} \Delta_{c} \bar{N} &=& \hat{n}^{i} D_{i} N 
  + 16 \pi G (V^{k} \hat{n}_{k}) \left[ \hat{n}^{i} \hat{n}^{j} P_{ij} - P/2
  \right] + \frac{d}{dt} (\sinh^{-1} \eta) - \Lie_{\bar{V}} (\sinh^{-1} \eta)
  \label{accel} \\
  \bar{\jmath}_{A} &=& 
  - 16 \pi G p_{A}^{\;\; i}  \tilde{P}_{ij} \hat{n}^j - d_{A} (\sinh^{-1} \eta)
  \label{j} \\
  \bar{N} \bar{s}^{AB} &=& - (16\pi G \hat{n}_{i} V^{i})
  \bar{\gamma}^{AC} \bar{\gamma}^{BD} \tilde{P}_{CD} +
  N ( k_{\hat{n}} \bar{\gamma}^{AB} - k_{\hat{n}}^{AB})
  + (\hat{n}^{i} D_{i} N) \bar{\gamma}^{AB}  \nn \\
  && + \bar{\gamma}^{AB} \left[ \frac{d}{dt} (\sinh^{-1} \eta ) 
  - \Lie_{\bar{V}} (\sinh^{-1} \eta) \right] \label{bars} 
\end{eqnarray}
With these expressions at hand, it is a reasonably straightforward exercise to
substitute them into (\ref{bulkss}).  We thereby obtain:
\begin{eqnarray} 
 & &  \int_{\Sigma}  \left\{ (\delta \bP^{ij} )(\delta_{\Lambda} h_{ij}) 
  - (\delta_{\Lambda} \bP^{ij})(\delta h_{ij}) \right\} \label{bulkssbdry} \\ 
  & = & \delta\left( H_{\Lambda}  
  - \int_{\mathcal{B}} \vBt \left\{ \frac{\lambda_{o}}{8\pi G} \bar{N}
    k_{\bar{n}} +  (\lambda_{o} \bar{V}^{A} + \lambda^{A}) \bar{\jmath}_{A}
    \right\} + \frac{1}{8 \pi G} \int_{S}\vS \tilde{\lambda}^{A} \tom_{A}
    \right) \nn \\ 
  && - \int_\Sigma \left\{ \delta(\lambda_o N)
    \bcH + \delta ( \lambda^i + \lambda_o V^i ) \bcH_i \right\} \nn \\ 
  && +\frac{1}{8 \pi G} \int_{S} \left\{ (\delta_{\Lambda} \vS)( \delta \ln
    \zeta)  - (\delta \vS)( \delta_{\Lambda} \ln \zeta ) \right\} 
    - \frac{1}{8 \pi G}\int_{S}(\delta \tilde{\lambda}^{A}) \vS \tom_{A} \nn \\ 
  && +\frac{\lambda_{o}}{8\pi G} \int_S  \left\{
    \frac{1}{2} \vS \tilde{s}^{AB} \delta \tilde{q}_{AB} 
    + \tilde{V}^{A} \delta(\tom_{A} \vS) 
    + \delta(\alpha C) \theta_{(n)} \vS 
    + (\delta \alpha) \vS \theta_{(\ell)} + 2 \delta( \alpha \vS \tl) 
    - \delta \left( \frac{d \vS}{dt} \right) \right\} \nn \\
  && + \frac{1}{8 \pi G} \int_{\mathcal{B}} 
    \left\{ (\delta_{\Lambda} \sinh^{-1} \eta ) (\delta \vBt) 
    - \delta (\sinh^{-1} \eta) (\delta_{\Lambda} \vBt) \right\} \nn \\
  && + \int_{\mathcal{B}} \vBt \left\{ \frac{\lambda_{o} \bar{N} }{16 \pi G} 
    \bar{s}^{AB}\delta \bar{\gamma}_{AB} 
    + \bar{\jmath}_{A} \delta(\lambda_{o} \bar{V}^{A}
    + \bar{\lambda}^{A}) + \frac{\delta( \bar{N} \lambda_{o}) }{8 \pi G}
    k_{\bar{n}} \right\} \, . \nn
    \end{eqnarray}
In the above, we have introduced the notation $H_{\Lambda}$ to signify the
bulk part of the Hamiltonian, namely
\begin{equation} 
  H_\Lambda = \int_\Sigma d^3 x \left\{ \lambda_o (N \bcH + V^i \bcH_i) 
  + \lambda^i \bcH_i  \right\} \, .  
\end{equation}
We can now interpret the terms arising in the expression (\ref{bulkssbdry})
one at a time.  The first term is an exact variation, which is precisely what
is required.  This will be part of the Hamiltonian which generates the
evolution along $\Lambda$.  The next term consists of variations of the lapse
and shift multiplying the bulk constraints.  This will clearly vanish provided
the bulk constraints are satisfied.  The remaining terms are all boundary
terms.  On each boundary, we have a contribution containing terms involving
the variation of the boundary volume forms $\vS$ and $\vBt$.  From these,
we identify the momenta conjugate to the volume forms as:
\begin{eqnarray}
  \tilde{P} =  \frac{1}{8 \pi G} \ln \zeta 
  \qquad \mbox{and} \qquad
  \bar{P} = \frac{1}{8 \pi G} \sinh^{-1} \eta \quad \, .
\end{eqnarray}
The equations of motion then guarantee that the evolution of these quantities
along $\Lambda$ is given by $\Lie_{\Lambda}$ as expected.

Finally we come to the remaining boundary terms at the horizon and outer
boundary.  In all cases, these terms either vanish (because $\tl = 0$) or the
quantity which appears inside variation is held ``fixed'' by our boundary
conditions.  Although these quantities are held fixed, with the weak boundary
conditions introduced in Section \ref{wbh}, their values will change under
diffeomorphisms of the boundary.  Thus, variations of these quantities do not
vanish, but are equal to Lie derivatives along a direction $X^i$ tangent to
the boundary surface.  For a three-dimensional Hamiltonian formulation this
means that
\begin{eqnarray}
  \delta_X = \delta t \left( \frac{\partial}{\partial t}  \right)
  + \Lie_{\widetilde{\delta X}} \, , 
\end{eqnarray}
as discussed in (\ref{X2}). 

In addition, we require that the diffeomorphism constraint be satisfied at
both the horizon and outer boundary.  The analysis of the outer boundary was
performed in detail in \cite{bf1}.  We refer the interested reader to that
paper (and in particular Appendix C); here we will simply use the result to
rewrite the terms arising in (\ref{bulkssbdry}).  Therefore, we turn our
attention to the horizon.  On $S$, the diffeomorphism constraint can be
rewritten in a form (\ref{prlCon}) which is directly applicable for our
current calculation (\ref{bulkssbdry}).  The constraint equation on the
horizon allows us to immediately cancel all the terms other than the
$\tilde{\omega}_{A}$ ones.  These can be rewritten using:
\begin{equation}
  \int_S \tilde{x}^A \Lie_{\mathcal{V}} ( \vS \tom_A) 
  = \int_S \left\{ \widetilde{\delta X}^A \frac{d}{dt} ( \vS \tom_A)
  + \tilde{V}^A \delta_X ( \vS \tom_A) \right\} \, ,
\end{equation}
Additionally, we must express the $\delta \tilde{\lambda}^{A}$ term in our
preferred co-ordinate system.  As in \cite{bf1}, we can make use of the
relation
\begin{equation}
   \delta \, \tilde{\lambda}^{\alpha} = 
   (\delta \tilde{\lambda})^{A}(d_{A}\tilde{\theta}^{\alpha}) 
   - \tilde{\lambda}^{A} d_{A} (\delta \tilde{\theta}^{\alpha}) \, , 
\end{equation}
to rewrite
\begin{eqnarray}
  \int_{S} (\delta\tilde{\lambda}^{A}) \vS \tom_{A} 
  + \lambda_{o} (\delta \tilde{\theta}^A ) \Lie_{T}(\vS \tilde{\omega}_{A}) 
  =  \int_{S} (\delta \tilde{\lambda}^{\alpha}) \vS  \tom_{\alpha} 
  + \delta \tilde{\theta}^{\alpha} \Lie_{\Lambda}(\vS \tilde{\omega}_{\alpha})
  \, . 
\end{eqnarray}
Finally combining everything from above, we can write the bulk symplectic
structure as:
\begin{eqnarray}
  \int_{\Sigma} && \left\{ (\delta \bP^{ij} )(\delta_{\Lambda} h_{ij}) 
  - (\delta_{\Lambda} \bP^{ij})(\delta h_{ij}) \right\} \label{bulkssfinal}  \\ 
  & = & \delta\left( H_{\Lambda}  
  - \int_{\mathcal{B}} \left\{ \frac{\lambda_{o}}{8\pi G} \vBt \bar{N}
    k_{\bar{n}} + \vBt (\lambda_{o} \bar{V}^{A} + \lambda^{A}) \bar{\jmath}_{A}
    \right\} + \frac{1}{8 \pi G} \int_{S}\vS \tilde{\lambda}^{A} \tom_{A}
    \right) \nn \\ 
  && - \int_\Sigma \left\{ \delta(\lambda_o N)
    \bcH + \delta ( \lambda^i + \lambda_o V^i ) \bcH_i \right\} \nn \\ 
  && + \int_{\mathcal{B}} \left\{ (\delta_{\Lambda} \bar{P}) (\delta \vBt) 
    - (\delta \bar{P} ) (\delta_{\Lambda} \vBt) \right\}
    + \int_{S} \left\{ (\delta_{\Lambda} \vS) (\delta \tilde{P})  
    - (\delta \vS)(\delta_{\Lambda} \tilde{P} ) \right\} \nn \\ 
  && + (\delta  \lambda_{o} ) \int_{\mathcal{B}}
    \left\{ \frac{ \bar{N}}{8 \pi G} \vBt k_{\bar{n}} 
    +\vBt \bar{\jmath}_{A} \bar{V}^{A} \right\} 
    + \delta t \int_{\mathcal{B}} \Lie_{\Lambda} \left\{ \frac{ \bar{N}}
    {8 \pi G} \vBt k_{\bar{n}} +\vBt \bar{\jmath}_{A} \bar{V}^{A} \right\} 
    \nn \\
  && + \int_{\mathcal{B}} \left\{ (\delta \bar{\lambda}^{\alpha}) \vBt 
    \bar{\jmath}_{\alpha} + \delta \bar{\theta}^{\alpha} \delta_{\Lambda}
    (\vBt \bar{\jmath}_{\alpha}) \right\}
    -\frac{1}{8 \pi G} \int_{S} \left\{ (\delta \tilde{\lambda}^{\alpha}) \vS 
    \tom_{\alpha} 
    + \delta \tilde{\theta}^{\alpha} \delta_{\Lambda}
    (\vS \tilde{\omega}_{\alpha}) \right\}  \, . \nn
\end{eqnarray}
This is the final desired form.

%%%%%%%%%%%%%%%%%%%%%%%%%%%%%%%%
%  REFERENCES
%%%%%%%%%%%%%%%%%%%%%%%%%%%%%%%%

\end{document}